\documentclass[preprint,12pt]{elsarticle}
\usepackage{amssymb}
\usepackage{amsmath}
\usepackage{graphicx}
\usepackage{lineno}
\usepackage{makecell} 
\usepackage{flushend}
\usepackage{longtable}
\usepackage{subcaption}
\usepackage{array}        
\usepackage{tabularx}     
\usepackage{url}
\usepackage{xurl}
\usepackage{makecell}     
\newcolumntype{Y}{>{\raggedright\arraybackslash}X} 
\usepackage{float}  
\journal{arXiv}

\begin{document}

\begin{frontmatter}

\title{Enhanced Prediction Model for Time Series Characterized by GARCH via Interval Type-2 Fuzzy Inference System} 

\author{Hongpei Shao}
\author{Da-Qing Zhang\corref{cor1}}
\author{Feilong Lu}

\cortext[cor1]{Corresponding author. Email: d.q.zhang@ustl.edu.cn} 

\affiliation{
	organization={Department of Mathematics, School of Science, University of Science and Technology Liaoning},
	addressline={189 Qianshan Middle Road, Lishan District},
	city={Anshan},
	postcode={114051},
	state={Liaoning Province},
	country={China}
}

\begin{abstract}
GARCH-type time series (characterized by Generalized Autoregressive Conditional Heteroskedasticity) exhibit pronounced volatility, autocorrelation, and heteroskedasticity. To address these challenges and enhance predictive accuracy, this study introduces a hybrid forecasting framework that integrates the Interval Type-2 Fuzzy Inference System (IT2FIS) with the GARCH model. Leveraging the interval-based uncertainty representation of IT2FIS and the volatility-capturing capability of GARCH, the proposed model effectively mitigates the adverse impact of heteroskedasticity on prediction reliability. Specifically, the GARCH component estimates conditional variance, which is subsequently incorporated into the Gaussian membership functions of IT2FIS. This integration transforms IT2FIS into an adaptive variable-parameter system, dynamically aligning with the time-varying volatility of the target series. Through systematic parameter optimization, the framework not only captures intricate volatility patterns but also accounts for heteroskedasticity and epistemic uncertainties during modeling, thereby improving both prediction precision and model robustness. Experimental validation employs diverse datasets, including air quality concentration, urban traffic flow, and energy consumption. Comparative analyses are conducted against models: the GARCH-Takagi-Sugeno-Kang (GARCH-TSK) model, fixed-variance time series models, the GARCH-Gated Recurrent Unit (GARCH-GRU), and Long Short-Term Memory (LSTM) networks. The results indicate that the proposed model achieves superior predictive performance across the majority of test scenarios in error metrics. These findings underscore the effectiveness of hybrid approaches in forecasting uncertainty for GARCH-type time series, highlighting their practical utility in real-world time series forecasting applications.
\end{abstract}

\begin{graphicalabstract}
	\includegraphics[width=0.8\textwidth]{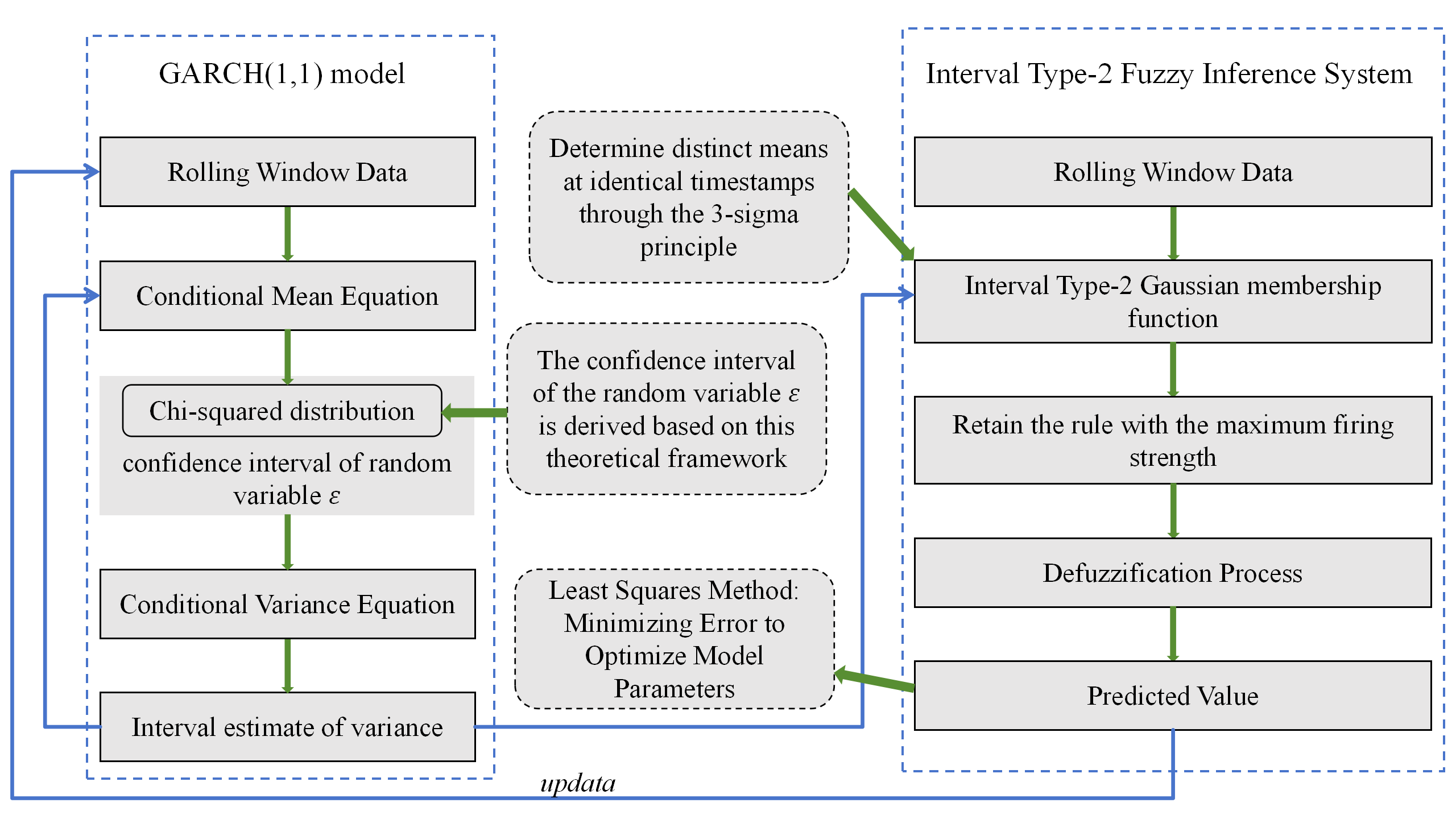}
\end{graphicalabstract}

\begin{highlights}
	\item The seamless integration of IT2FIS with GARCH reduces the impact of heteroskedasticity and uncertainty in volatile time series on model predictions.
	\item The GARCH-estimated conditional variance is embedded into the Gaussian membership functions of IT2FIS, transforming IT2FIS into an adaptive fuzzy inference system that dynamically aligns with real-time volatility.
	\item The defuzzification mechanism of IT2FIS-GARCH refines the conditional mean prediction of the GARCH model.
\end{highlights}

\begin{keyword}
Generalized Conditional Heteroskedasticity Model, Fuzzy Inference System, Interval Type-2 Membership Function, Adaptive Parameter Prediction Model

\end{keyword}

\end{frontmatter}

\section{Introduction}
\label{Introduction}
GARCH-type time series (time series characterized by Generalized Autoregressive Conditional Heteroskedasticity) are widely recognized for their inherent heteroskedasticity, a phenomenon where variance exhibits significant temporal variability. Such volatility clustering—alternating phases of high and low dispersion—is particularly prevalent in financial markets, manifesting in stock prices, exchange rates, and interest rate fluctuations\cite{yolcu_novel_2023, jiang_grey_2012}. 

Methodologically, time series forecasting frameworks are broadly categorized into linear and nonlinear paradigms. Linear models, such as the Autoregressive Moving Average (ARMA), Autoregressive Integrated Moving Average (ARIMA), and their heteroskedasticity-aware extensions—the Autoregressive Conditional Heteroskedasticity (ARCH) \cite{engle_autoregressive_1982} and Generalized Autoregressive Conditional Heteroskedasticity (GARCH) \cite{bollerslev_generalized_1986} models—leverage historical linear dependencies for prediction. For instance, adaptive feature fusion mechanisms integrated with ARIMA have enhanced remaining useful life predictions for high-speed train bearings \cite{guo_adaptive_2024}, achieving significant improvements in forecasting accuracy. In the renewable energy domain, \citet{chen_wind_2018} effectively combined an outlier-smoothed transition autoregressive structure with the GARCH model, enabling efficient prediction of wind power volatility. Beyond hybrid modeling approaches for performance enhancement, forecasting precision can also be improved through model parameter optimization. Hybrid methodologies further augment performance; for example, the Grey Wolf Optimizer (GWO) algorithm has been employed to optimize GARCH and ARIMA parameters, yielding higher accuracy in stock price predictions \cite{bagalkot_novel_2024}. Despite these advancements, linear models face inherent limitations in capturing complex nonlinear dependencies, primarily due to their insufficient capacity to characterize nonlinear patterns, which may lead to degraded forecasting performance. Consequently, researchers increasingly emphasize the integration of nonlinear frameworks or hybrid architectures to address such challenges. Building on this rationale, this study introduces a novel hybrid model that synergizes the Interval Type-2 Fuzzy Inference System (IT2FIS) with GARCH, aiming to mitigate heteroskedasticity-induced uncertainties while enhancing adaptability to nonlinear patterns.

Nonlinear models provide flexible and versatile approaches for time series forecasting. Mainstream nonlinear methodologies in this domain include neural networks, support vector machines (SVMs), random forests (RF), and gradient boosting machines (GBMs). For example, an integration of artificial neural networks (ANNs) with autoregressive integrated moving average (ARIMA) models was proposed in \cite{senol_estimation_2025} to predict biogas potential from poultry manure in Turkey, demonstrating the capability of neural networks to capture nonlinear data characteristics through multilayer architectures and nonlinear activation functions. Similarly, a physics-informed neural network framework for fatigue life prediction was developed in \cite{wang_multi-physics_2024}, where fatigue limits were embedded into activation functions and a hybrid loss function incorporating physical laws was formulated. Beyond neural networks, machine learning techniques such as SVM and RF have also exhibited significant advantages in time series forecasting. An improved random forest-based method for ultra-short-term photovoltaic cluster output power prediction was introduced in \cite{yang_improved_2021}, addressing low accuracy and overfitting issues in conventional models. Additionally, an SVM model enhanced by principal component analysis and the Shapley swarm algorithm was designed in \cite{tong_guangxi_2024} for high-precision GDP forecasting. These studies collectively highlight the superior ability of nonlinear models to characterize complex temporal patterns. However, a critical limitation of nonlinear models—particularly neural networks—is their dependency on large-scale training datasets and meticulous hyperparameter tuning to effectively learn inherent data patterns, which poses challenges for practical implementation.

The Interval Type-2 Fuzzy Inference System (IT2FIS), as a nonlinear modeling tool, demonstrates superior performance in handling uncertainties through the integration of fuzzy sets and inference rules. Its strong robustness against external disturbances, noise, and measurement uncertainties has led to widespread adoption across diverse domains \cite{celikyilmaz_uncertainty_2008, melin_improved_2010, castillo_design_2011}. Compared to neural networks, IT2FIS offers distinct advantages, including fewer tunable parameters and a more interpretable architecture, thereby providing an efficient solution for modeling complex systems. For instance, \citet{zhangNovelHybridDeep2024} integrated Type-1 and interval Type-2 Takagi-Sugeno-Kang (TSK) fuzzy modules, significantly improving prediction accuracy and generalization capability through fuzzy parameter optimization. In time series forecasting, IT2FIS exhibits notable adaptability and application potential. \citet{das_evolving_2015} proposed a meta-cognitive interval Type-2 neuro-fuzzy inference system (McIT2FIS) with integrated learning mechanisms, demonstrating superior performance over conventional fuzzy systems in time series prediction and stock price tracking. Furthermore, the fusion of IT2FIS with optimization algorithms has expanded its applicability. For example, \citet{zou_inter_2019} developed a fuzzy interval prediction model based on an interval Type-2 framework optimized via a gravitational search algorithm, achieving significant accuracy improvements. Collectively, these advancements underscore IT2FIS's pivotal role in time series forecasting, with its exceptional uncertainty-handling capabilities driving widespread academic recognition and practical adoption since its inception.

Despite these advantages, conventional Interval Type-2 Fuzzy Inference Systems (IT2FIS) exhibit limitations in modeling GARCH-type time series due to their static parameterization. Traditional implementations fix membership function parameters during training and retain them throughout inference, which fails to capture the dynamic conditional heteroskedasticity inherent in such series—a critical gap given the time-varying volatility patterns of GARCH processes. To address this, we propose the IT2FIS-GARCH framework, a novel hybrid architecture that dynamically embeds GARCH-predicted variance estimates into Gaussian membership functions. This integration enables adaptive parameter adjustments aligned with observed volatility dynamics, bridging the methodological divide between fuzzy logic and econometric volatility modeling.

Within the IT2FIS-GARCH framework, Gaussian membership function parameters are no longer static but evolve temporally based on GARCH-derived variance signals. At each forecasting step, these parameters are optimized to reflect real-time heteroscedastic patterns, transforming IT2FIS into a variable-parameter system. Specifically, the variance estimates drive the adaptive tuning of fuzzy inference rules, allowing the model to simulate the volatility clustering behavior characteristic of GARCH-type series. This mechanism not only mitigates the rigidity of traditional IT2FIS but also enhances predictive accuracy by  time-varying uncertainty structures. Crucially, the proposed approach preserves IT2FIS's interpretability, offering a unified solution for complex forecasting tasks.

A critical limitation of traditional GARCH models lies in their conditional mean equation, which assumes a constant mean term \cite{lin_modelling_2018}. While this assumption holds for stationary processes with stable mean levels, it becomes invalid when modeling heteroscedastic time series exhibiting time-varying mean dynamics—a common scenario in financial and energy markets. To overcome this constraint, the proposed IT2FIS-GARCH model explicitly incorporates conditional heteroskedasticity-aware mechanisms into the mean prediction process through adaptive defuzzification. Specifically, the system dynamically adjusts output centroids based on real-time volatility estimates derived from GARCH, enabling joint optimization of mean and variance predictions. This dual-target adaptation effectively captures both non-stationary trends and volatility clusters, addressing the rigidity of conventional GARCH specifications.

To evaluate model performance, the proposed model is benchmarked against the following models: the Generalized Autoregressive Conditional Heteroskedasticity-Takagi-Sugeno-Kang (GARCH-TSK) model, Fixed Variance Time Series models, GARCH-Gated Recurrent Unit (GARCH-GRU), and Long Short-Term Memory (LSTM) networks. Experimental results demonstrate , the proposed model significant reductions in error metrics and superior $R^2$ values compared to other models multiple datasets, validating its enhanced predictive accuracy and robustness.

\section{Related Work}
\label{Related Work}
In recent years, heteroskedastic time series forecasting has garnered wides-\\pread attention and research. The GARCH model, as a significant econometric model, is specifically designed to characterize the volatility features of time series data, with its core focus on describing the non-constancy of variance. The GARCH model was developed on the basis of the ARCH model, aiming to overcome the limitation of the ARCH model, which can only effectively capture short-term volatility clustering effects. This model, grounded in the theory of conditional heteroskedasticity, predicts future volatility by incorporating historical volatility information. From a theoretical framework perspective, the GARCH model captures the volatility characteristics of time series by expressing the conditional variance as a linear combination of past squared residuals and past conditional variances. The simplest form of the GARCH model is GARCH(1,1):

\begin{equation}
	\begin{aligned}
		Y _ { t } &= \mu + \varepsilon _ { t }\sigma _ { t }, \\
		\varepsilon _ { t } &\sim N(0,1),\\
		\sigma _ { t } ^ { 2 } &= w + \alpha _ { 1 } \varepsilon _ { t - 1 } ^ { 2 }\sigma_{ t - 1 }^2 + \beta _ { 1 } \sigma _ { t - 1 } ^ { 2 }.
	\end{aligned}
	\label{eq:GARCH}
\end{equation}

Where $Y_{t} $ is the data value at time $t $, $ \mu $ is the conditional mean, $ \varepsilon $ is usually assumed to follow a normal distribution, $ \sigma_t ^ 2 $ is the conditional variance of the input data at time $t $, $ w $ is a constant term in the model, $ \alpha_1 $ is used to represent the impact of the square of past residuals on the current volatility, $ \beta_1 $ is used to represent the impact of the square of past volatility on current volatility. In the formula \eqref {eq:GARCH}, $Y_t$ is the conditional mean equation, $ \sigma_{t} ^ {2} $ is the conditional variance equation.

In the study of heteroscedastic time series, the GARCH model and the ARMA model serve different modeling functions: the GARCH model is primarily used to characterize the volatility characteristics of the time series, while the ARMA model focuses on describing the linear structural characteristics of the data. To fully leverage the advantages of both models, researchers have proposed the idea of combining them. For example, in the literature \cite{liuSPIbasedDroughtSimulation2019a}, the GARCH model is introduced into the residual sequence of the ARMA model, effectively eliminating the ARCH effect in the residuals and constructing an ARMA-GARCH composite model. In the literature \cite{ekinciModellingForecastingGrowth2021}, an ARMA model is embedded in the mean equation of the GARCH model. In this framework, the ARMA component is mainly used to capture the dynamic changes of the time series itself, while the GARCH component focuses on describing the volatility dynamics of the residual sequence. This dual-component architecture preserves ARMA’s capacity to represent linear trends while leveraging GARCH’s ability to model volatility clustering. By decomposing time series into distinct mean and variance processes, hybrid models achieve comprehensive characterization through specialized submodels, significantly improving forecasting accuracy. These integrative approaches establish rigorous theoretical foundations and provide actionable methodologies for complex heteroscedastic series analysis.

In addition, the fuzzy GARCH model, as an important extended model, has gained widespread attention in recent years. For instance, \citet{hung_applying_2011} pioneered a genetic algorithm (GA)-driven fuzzy GARCH model, where membership functions and GARCH parameters are co-optimized via iterative evolutionary search, achieving marked improvements in volatility forecasting precision. Building on this, \citet{hungAdaptiveFuzzyGARCHModel2011a} developed an adaptive variant leveraging particle swarm optimization (PSO) to address nonlinearities in equity market volatility, demonstrating superior convergence properties over gradient-based methods. Further advancing this domain, \citet{almeidaEstimationFlexibleFuzzy2014a} introduced a hybrid architecture that synergizes probabilistic GARCH variance modeling with fuzzy rule-based linguistic uncertainty handling, enabling multidimensional characterization of heteroscedastic dynamics. Crucially, these frameworks embed GARCH-derived variance calculations into the fuzzy rules layers of fuzzy inference systems, applying fuzzy-weighted correction mechanisms to refine volatility estimates. Collectively, fuzzy GARCH models unify the interpretability of fuzzy logic with the statistical rigor of econometric volatility modeling, offering a versatile toolkit for complex time series forecasting—a convergence that bridges epistemic uncertainties with data-driven variance dynamics.

To address existing research gaps, this study proposes a novel hybrid framework—the IT2FIS-GARCH model—that synergistically integrates conditional heteroskedasticity modeling with interval type-2 fuzzy inference systems (IT2FIS). The key innovations are threefold:

1. Dynamic Volatility Adaptation: By embedding the conditional variance estimated by the GARCH model into the Gaussian membership functions of IT2FIS, our framework transforms IT2FIS into an adaptive fuzzy inference system. This enables real-time alignment with the volatility of the target series.

2. Mean-Variance Co-Optimization: The defuzzification mechanism based on the IT2FIS-GARCH model optimizes the conditional mean prediction of the GARCH model, jointly optimizing trend and forecasts.

The structure of this paper is organized as follows: Section 3 elaborates on the modeling process and methodological framework of the IT2FIS-GARCH model; Section 4 details the specific forecasting procedure of the proposed model; Section 5 presents a comparative analysis of different models' forecasting performance to validate the effectiveness of the IT2FIS-GARCH model; finally, Section 6 concludes the paper with a summary of key findings.

\section{The Establishment of the IT2FIS-GARCH Model}
\label{The establishment of the IT2FIS-GARCH model}
IT2FIS (Interval Type-2 Fuzzy Inference Systems), a significant advancement in the domain of fuzzy inference systems, represents an evolved form of traditional Type-1 Fuzzy Inference Systems (T1FIS). The incorporation of interval-valued membership functions grants IT2FIS enhanced flexibility and adaptability in managing uncertainties.

From a structural perspective, one of the core aspects of IT2FIS is reflected in the construction of its membership functions. Unlike T1FIS, which relies on deterministic membership functions, IT2FIS utilizes interval-based membership functions, where each input value maps to a bounded membership range rather than a single crisp value. In practical applications, commonly used membership functions include Gaussian, trapezoidal, Z-shaped, and S-shaped membership functions. Taking the interval-type Gaussian membership function as an example, it can be represented in IT2FIS as:$\mu(x) = \left\{ \mu(x; c, \sigma) \middle| \sigma \in [\sigma_1, \sigma_2] \right\}$
. Where $\mu(x; c, \sigma)$ represents the Gaussian membership value under the given $c$ and $\sigma$, with $\sigma_1 \leqslant \sigma \leqslant \sigma_2$. Therefore, $\mu(x)$ is an interval value, indicating that the membership degree of element $x$ to the fuzzy set varies within a range.

The rule base of IT2FIS stores a series of fuzzy rules used for reasoning, and the rule structure can be expressed as:
\begin{equation}
	\begin{aligned}
		&\text{IF } x_1 \text{ is } A_1 \text{ AND } x_2 \text{ is } A_2 \ldots \text{ AND } x_n \text{ is } A_n\\
		&\text{THEN } y \text{ is } a_{0} + \sum_{j=1}^{t} a_{j} x_{j}
	\end{aligned}
	\label{eq:if then}
\end{equation}
Here, $x_1,\ldots,x_n$ are the input variables, $A_1, \ldots, A_n$ are the type-2 fuzzy sets of the input variables, and $y$ is the output variable. It is important to note that the rule consequent represents the relationship or mapping between the output variable $y$ and the input variable set $x_1,\ldots,x_n$.

The inference mechanism of IT2FIS performs reasoning based on input data and the rules in the rule base, then generates outputs.

Therefore, the model studied in this paper is referred to as the IT2FIS-GARCH model, and the subsequent sections will detail its construction process and implementation methods. Below is a brief description of the mathematical principles of the model:
\begin{equation}
	[\underline{\mu}_t, \overline{\mu}_t] = mf\left([\underline{\sigma}_{t}, \overline{\sigma}_{t}], \ldots, [\underline{\sigma}_{t-k+1}, \overline{\sigma}_{t-k+1}]\right), 
	\label{eq:IT2FIS-GARCH-mf}
\end{equation}

\begin{equation}
	\hat{Y}_{t} = IT2FIS\left(input_{t}, [\underline{\mu}_t, \overline{\mu}_t] \right).
	\label{eq:IT2FIS-GARCH}
\end{equation}

\begin{equation}
	Y_{t} = \hat{Y}_{t} + \varepsilon_{t}\sigma_{t}, \quad \varepsilon_{t} \sim \mathcal{N}(0,1).
	\label{eq:junzhifangcheng}
\end{equation}

\begin{equation}
	\sigma_{t}^2 = \omega + \alpha_{1} \varepsilon_{t-1}^2\sigma_{t-1}^2 + \beta_{1}\sigma_{t-1}^2.
	\label{eq:tiaojianyifangchafangcheng}
\end{equation}
Here, IT2FIS is an interval type-2 fuzzy inference system. $input_{t}$ represents the data window input to IT2FIS at time $t$, and this window contains $k$ consecutive data points. $[\underline{\mu}_{t}, \overline{\mu}_{t}]$ is the interval Gaussian membership value. $mf$ represents the membership function of IT2FIS. $[\underline{\sigma}_{t}, \overline{\sigma}_{t}], \ldots, [\underline{\sigma}_{t-k+1}, \overline{\sigma}_{t-k+1}]$ are the interval variances corresponding to the $k$ data points, with each data point having a different variance. The variation in these variances is no longer simply treated as a static parameter, but is dynamically represented in the Gaussian membership function of IT2FIS.

In terms of overall architecture, IT2FIS is capable of targeted processing of complex nonlinear relationships and uncertain information presented in time series based on the characteristics of interval type membership functions and complex rule libraries. The organization of Section \ref{The establishment of the IT2FIS-GARCH model} in this paper is as follows: subsection \ref{Training and optimization of the GARCH model} focuses on the GARCH model, detailing its training process and the methods used for parameter estimation. Subsection \ref{Estimation of GARCH model parameters} centers on the time-varying parameter IT2FIS, emphasizing its structural design principles and the specific procedure for rule establishment. Additionally, to facilitate readers' understanding and practical implementation, algorithm 3.1 provides the detailed code for the IT2FIS-GARCH model during the modeling process.

\subsection{Training and optimization of the GARCH model}
\label{Training and optimization of the GARCH model}
\begin{figure}[H]
	\centering
	\includegraphics[width=0.8\textwidth]{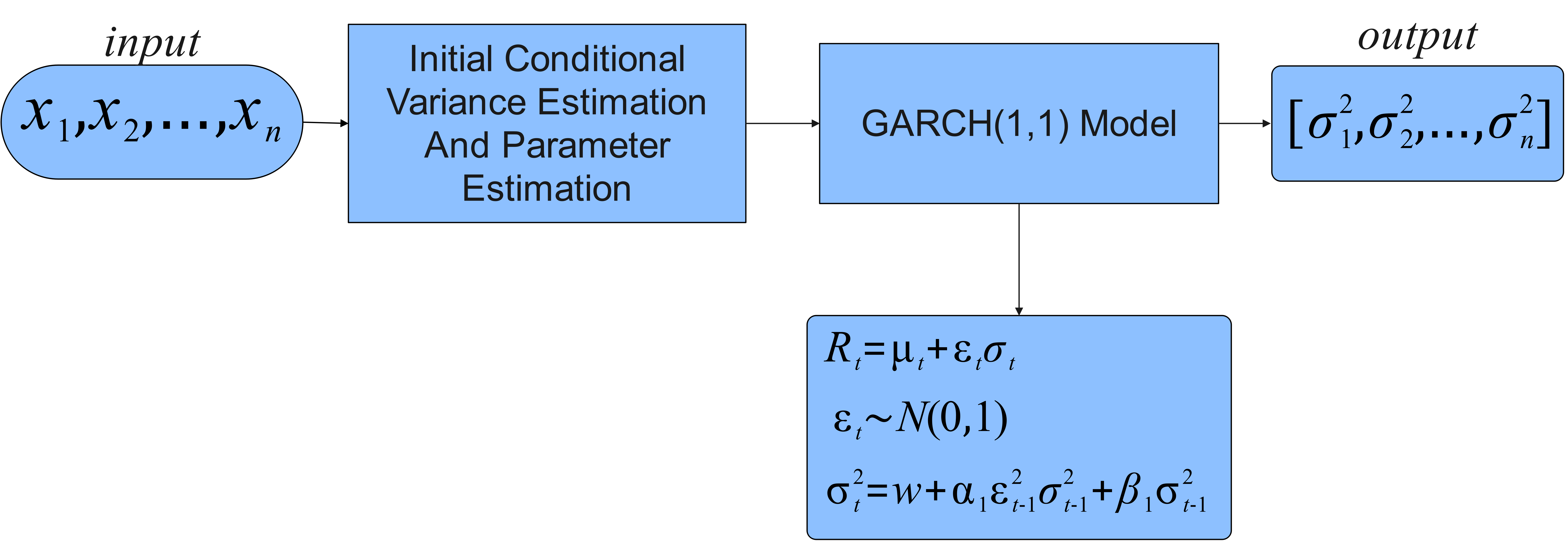}
	\caption{GARCH(1,1) model. For a dataset with $n$ data points, the process is as follows: First, determine the initial conditional variance. Then, estimate the model parameters. Finally, use an iterative algorithm to optimize the parameters $w$, $\alpha_1$ and $\beta_{1}$, and implement the prediction for the GARCH(1,1) model.}
	\label{fig:GARCH Model}
\end{figure}

Equation \eqref{eq:GARCH} presents the general form of the GARCH model. The parameters of the model, $w$, $\alpha_{1}$, and $\beta_{1}$, are obtained by fitting the GARCH model. The workflow of the GARCH model is shown in Figure \ref{fig:GARCH Model}.

During the training process of the GARCH(1,1) model, Equation \eqref{eq:GARCH} demonstrates that the GARCH(1,1) model primarily focuses on point prediction of the time series variance. However, for forecasting heteroskedastic time series, a single point prediction approach struggles to fully capture its heteroskedastic nature. To more accurately characterize the volatility dynamics of the time series variance, adopting an interval prediction method to estimate the range of variance fluctuations proves to be a more reasonable and effective strategy. The procedure for variance interval estimation is presented in Subsubsection \ref{Interval estimation of variance}.

\subsubsection{Interval estimation of variance}
\label{Interval estimation of variance}
From equation \eqref{eq:GARCH}, it is known that the random variable $\varepsilon_t$ follows a normal distribution with a mean of 0 and a variance of 1. According to probability theory, $\varepsilon_t^2$ follows a chi-square distribution with degrees of freedom $df=1$, denoted as $\varepsilon_t^2 \sim \chi^2(1)$. Based on this distribution property, in the prediction of heteroscedastic time series, the confidence interval for $\varepsilon_t^2$ can be obtained by referring to the chi-square distribution table, which can then be used to derive the interval estimate of the conditional variance. First, the confidence interval for $\varepsilon_t^2$ at a confidence level of ($1 - \alpha$) is as follows:
\begin{equation}
	\left[ \chi^2_{1, \alpha/2} ,\chi^2_{1, 1-\alpha/2} \right]
	\label{eq:chi}
\end{equation}
Where $ \alpha $ is the significance level, $\chi^2_{1, \alpha/2}$ and $\chi^2_{1, 1-\alpha/2}$ are the lower $\alpha/2$ and upper $1-\alpha/2$ quantiles of the chi-square distribution with 1 degree of freedom, respectively. After obtaining the confidence interval at the significance level, the interval estimate of the variance can be derived based on the conditional heteroskedasticity equation in formula \eqref{eq:GARCH}. The specific formula is as follows:
\begin{equation}
	\begin{aligned}
		\overline{\sigma_t^{2}} &= w + \alpha _ { 1 } \chi_{df, 1-\alpha/2} ^ { 2 } \sigma_{t-1}^2+ \beta _ { 1 } \sigma _ { t - 1 } ^ { 2 },\\
		\underline{\sigma_t^{2}} &= w + \alpha _ { 1 } \chi_{df, 1-\alpha/2} ^ { 2 } \sigma_{t-1}^2+ \beta _ { 1 } \sigma _ { t - 1 } ^ { 2 }.
	\end{aligned}
	\label{eq:qujian}
\end{equation}

\begin{figure}[H]
	\centering
	\includegraphics[width=0.8\textwidth]{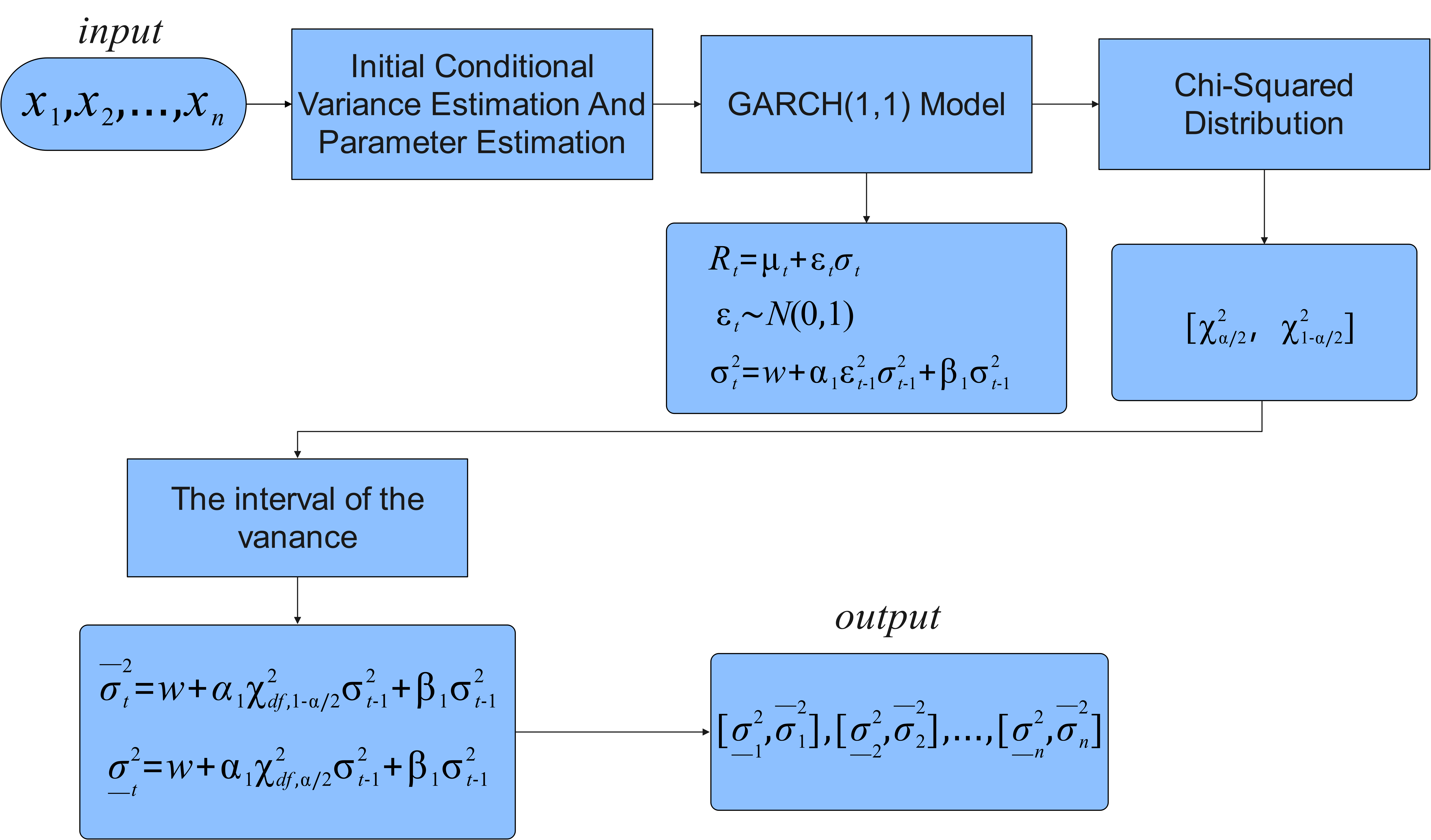}
	\caption{Variance interval estimation based on GARCH model. For a dataset with $n$ data points, obtain interval estimates of the variance for each data point. Firstly, estimate the initial variance value and model parameters. Then, refer to the chi square distribution table to obtain $\chi^2_{\alpha/2}$ and $\chi^2_{1 - \alpha/2}$. Finally, use the interval values of the random variables to calculate the variance interval values of the data. This leads to the final implementation of interval estimation of variance.}
	\label{fig:Confidence Interval for Variance}
\end{figure}

The flowchart for variance interval estimation based on the GARCH model is illustrated in Figure \ref{fig:Confidence Interval for Variance}.

\subsection{Estimation of GARCH model parameters}
\label{Estimation of GARCH model parameters}

As described in Subsubsection \ref{Interval estimation of variance}, this study employs the GARCH model to estimate the interval values of variance, the effectiveness of which relies on precise parameter identification. This section elaborates in detail on the training process of GARCH model parameters.

Suppose there is a known dataset $X =\{x_1,  x_2, \cdots, x_n\}$ consisting of $n$ data points. Assume the sliding window size is 5, then the input data at time $t$ is $[x_{t-4}, \cdots, x_t]$. Initially, assume the initial variance $\sigma_{0}=0$, and the initial parameter values are $\{w, \alpha,  \beta\}=\{1, 1, 1\}$. The initial residual term $\varepsilon _ { 0 }$ is a random number following a normal distribution. Based on the variance interval estimation method described in Subsubsection \ref{Interval estimation of variance}, the variance at time $t+1$ can be expressed as:
\begin{equation}
	[\underline{\sigma^2}_{t+1},\overline{\sigma^2}_{t+1}]
	\label{eq:sigma1}
\end{equation}

The obtained interval variance is input into the Interval Type-2 Fuzzy Inference System, which is used to predict the GARCH-type time series and obtain the corresponding predicted value $\hat{Y}_{t+1}$. For detailed steps and implementation methods of the fuzzy inference process, please refer to Section \ref{IT2FIS-GARCH model} of this paper. According to formula \eqref{eq:junzhifangcheng}, update the residual term $\varepsilon$ at time $t+1$. Therefore, the update method of $\varepsilon$ can be expressed as:
\begin{equation}
	\varepsilon_{t+1} = \frac { Y _ { t+1  } - \hat{Y} _ { t+1  } } { \sigma _ { t+1 } }
	\label{eq:varepsilon}
\end{equation}

After obtaining $\varepsilon_{t+1}$ from the predicted value $\hat{Y}_{t+1}$ and formula \eqref{eq:junzhifangcheng}, the variance for the next time step is obtained again using formulas \eqref{eq:chi} and \eqref{eq:qujian}. By analogy, continuously utilizing new observations and predicted variances to obtain new residual terms. In this way, the variance prediction values for each time step in the time series can be calculated sequentially, and the corresponding residual terms for each time step can be obtained. The initial parameters are set to $\{w, \alpha, \beta\}=\{1, 1, 1\}$, and the mean square error (MSE) between the predicted value $\hat{Y} $ and the true value $Y $ is calculated using the least squares method to iteratively optimize these parameters. This process aims to identify the optimal parameter combination $\{ w_ {beat}, \alpha_ {best}, \beta_ {best} \} $, and the mean square error calculation formula is as follows:
\begin{equation}
	\text{MSE} = \frac{1}{n} \sum_{i=1}^{n} (Y_i - \hat{Y}_i)^2
	\label{eq:mse}
\end{equation}
Here, $ n $ represents the sample size, $Y_i$ represents the observed actual value, and $\hat{Y}_i$ represents the predicted value of the model.
\begin{table}[H]
	\centering
	
	\footnotesize
	\begin{tabular}{l}
		\hline
		\textbf{Algorithm 3.1} IT2FIS-GARCH: Modeling code for the IT2FIS-GARCH model. \\\hline

		\textbf{Input:}Training set $\tilde { X }_{train} = \{x_1, x_2, \ldots,  x_i\}$,\\
		initialization $w, \alpha, \beta$, initial rule consequent parameters $a_0, a_1, \ldots, a_t$.\\
		\textbf{Output:}Optimal coefficient value $w^{best}, \alpha^{best}, \beta^{best},a_0^{best},a_1^{best}, \ldots, a_t^{best}$.\\
		1.Replace the outliers in $\tilde {X} _ {train} $ using the linear difference method;\\
		
		2.Set the initial parameters to 1;\\
		3.The starting $\varepsilon_0 $ is a random number that follows a standard normal distribution;\\
		4.Obtain the mean list $mean^{act}$;\\
		5.Sliding window size $W$;\\
		\textbf{for} $t=1$ \textbf{to} $len(\tilde { X }_{train})$\\
		\hspace{2em} Add the current value $x_ {t} $ to the window $win $;\\
		\hspace{2em} \textbf{if} $len(win)\textgreater W$ \textbf{then}\\
		\hspace{4em} Delete the oldest data in the window;\\
		\hspace{2em} \textbf{end if}\\
		\hspace{2em} Calculate the interval value of $\varepsilon ^ 2 $ based on\\
		\hspace{2em} the chi square distribution table;\\
		\hspace{2em} Predict the variance at time $t $ by $\sigma_{t}^2 = w + \alpha \varepsilon_{t-1}^2\sigma_{t-1}^2 + \beta \sigma_{t-1}^2$;\\
		\hspace{2em} Obtain the interval value of variance [$\underline{\sigma}_{t}^2$, $\overline{\sigma}_{t}^2$];\\
		\hspace{2em} Calculate the mean of the data based on the $3 \ sigma $ criterion;\\
		\hspace{2em} Calculate the interval membership value of data based on\\
		\hspace{2em} Gaussian membership function\\
		\hspace{2em} $[\underline{\mu} _ { t },\overline{\mu} _ { t }]$;\\
		\hspace{2em} Match fuzzy rules and output the final predicted value $\hat{Y}_t = a_{0} + \sum_{j=1}^{t} a_{j} x_{j}$,\\
		\hspace{2em} Store predicted values in the predicted value list $\hat{Y}$;\\
		\hspace{2em} Calculate $\varepsilon$ at time $t $;\\
		\hspace{2em} Update $\sigma_t^2$ and $\varepsilon_t^2$;\\
		\textbf{end for}\\
		$\hat{Y}$ \textbf{is obtained}\\
		Calculate the error value $e $ between the predicted value and the actual value;\\
		
		Set the maximum number of iterations $n $;\\
		\textbf{for} $1$ \textbf{to} $n$\\
		\hspace{2em} Generate new model parameters;\\
		\hspace{2em} Generate predicted values for new parameters based on\\
		\hspace{2em} the code from the previous loop;\\
		\hspace{2em} Calculate the error value $e_ {new} $ under the new parameters;\\
		\hspace{2em} \textbf{if} $e_{new}$ \textgreater $e$ \textbf{then}\\
		\hspace{4em} Update parameters;\\
		\hspace{4em} Update error values;\\
		\hspace{2em} \textbf{end if}\\
		\textbf{end for}\\
		\textbf{return} $w^{best},\alpha^{best},\beta^{best},a_0^{best},a_1^{best},\ldots,a_t^{best}$\\
		
		\hline
	\end{tabular}	
\end{table}

\subsection{Establishment and parameter estimation of the time-varying parameter IT2FIS}
\label{the Time-Varying Parameter IT2FIS}
\label{Establishment and Parameter Estimation of the Time-Varying Parameter IT2FIS}
\begin{figure}[H]
	\centering
	\includegraphics[width=0.8\textwidth]{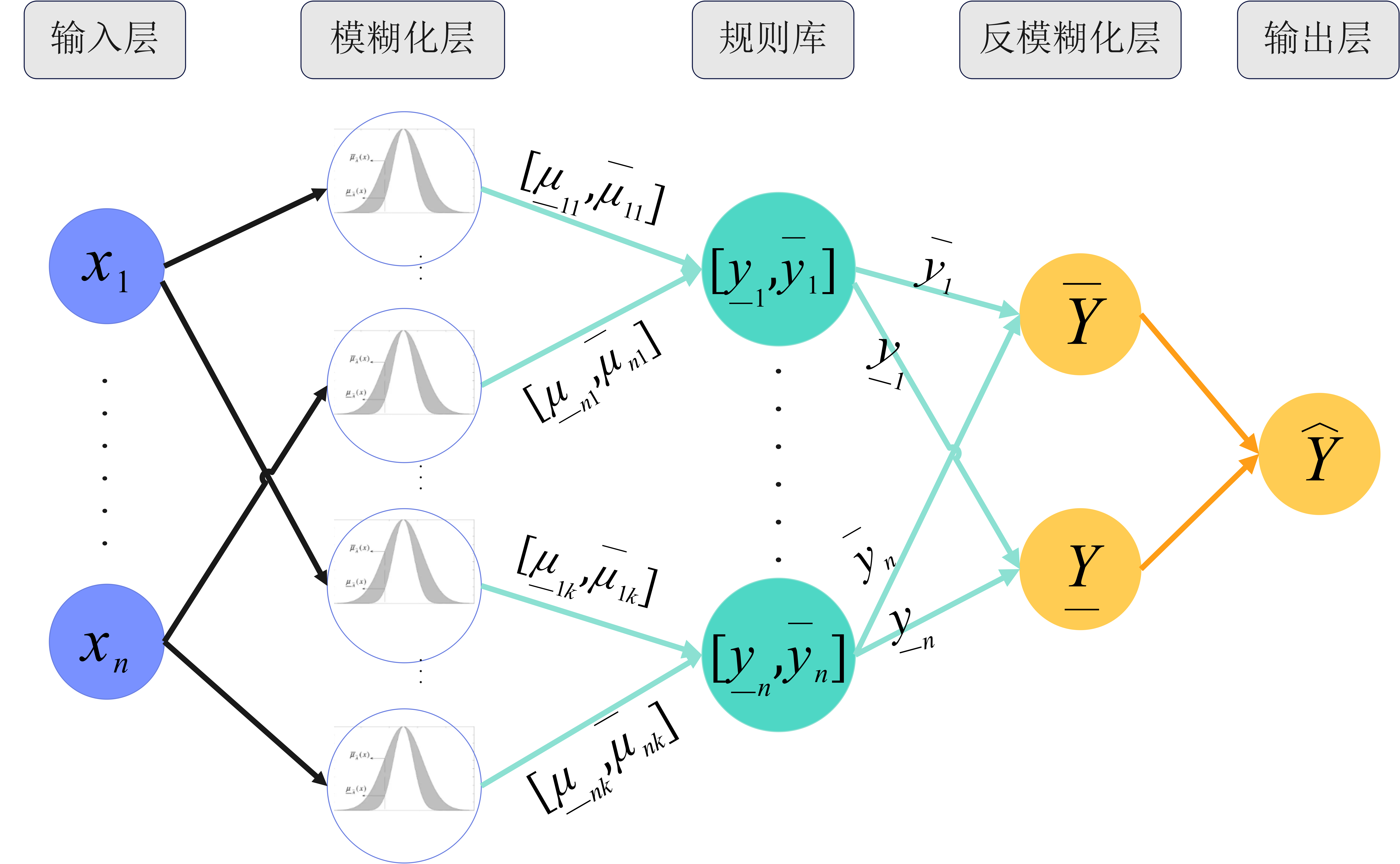} 
	\caption{Interval type-2 fuzzy reasoning system. For a dataset with $n$ data points, input it into the second layer of the system and use the membership function to calculate the membership interval $[\underline{\mu}_{nk},\overline{\mu}_{nk}]$. Taking Gaussian membership function as an example in the figure. Match rules based on the rule library and output the corresponding rule consequent $[\underline{y}_n,\overline{y}_n]$. Defuzzify the fuzzy output result to obtain a clear value $\hat{Y}$, which serves as the final output of the system.}
	\label{fig:TT2FIS}
\end{figure}

The working process of the traditional IT2FIS is shown in Figure\ref{fig:TT2FIS}, where $y_i(i=1,...,n)$ represents the output of the rule consequent, $Y$ represents the fuzzy output of the model, and $\hat{Y}$ represents the predicted value. The model consists of four parts: the input layer, fuzzification layer, rule base, defuzzification layer, and output layer.

\subsubsection{Variable parameter gaussian membership function}
\label{Gaussian membership function}
In the second layer of the time-varying parameter IT2FIS, membership functions are employed to characterize the degree of membership between input values and fuzzy sets. Among various membership functions, the Gaussian membership function has been widely adopted due to its smoothness and superior adaptability to uncertainty and fuzzy problems. Furthermore, the Gaussian membership function demonstrates certain adaptive capabilities to heteroskedasticity in specific scenarios. Based on these advantages, this study utilizes Gaussian membership functions to model heteroskedasticity in time series. The Gaussian membership function in the time-varying parameter IT2FIS incorporates two critical parameters: the interval variance and mean, whose appropriate determination is crucial for model performance.

A gaussian fuzzy set of an IT2FIS, denoted as $\tilde {A}$, has a defined mean $m$, an upper standard deviation $ \overline{ \sigma } $ and a lower standard deviation $ \underline{\sigma} $, where $\tilde {A}$ is represented as $\tilde {A} = <m, [\underline{ \sigma },\overline{\sigma}]>$, and $ \underline{ \sigma } \neq \overline{\sigma} $ \cite{chen_fuzzy_2013}.

\begin{figure}[H]
	\centering
	\includegraphics[width=0.9\textwidth]{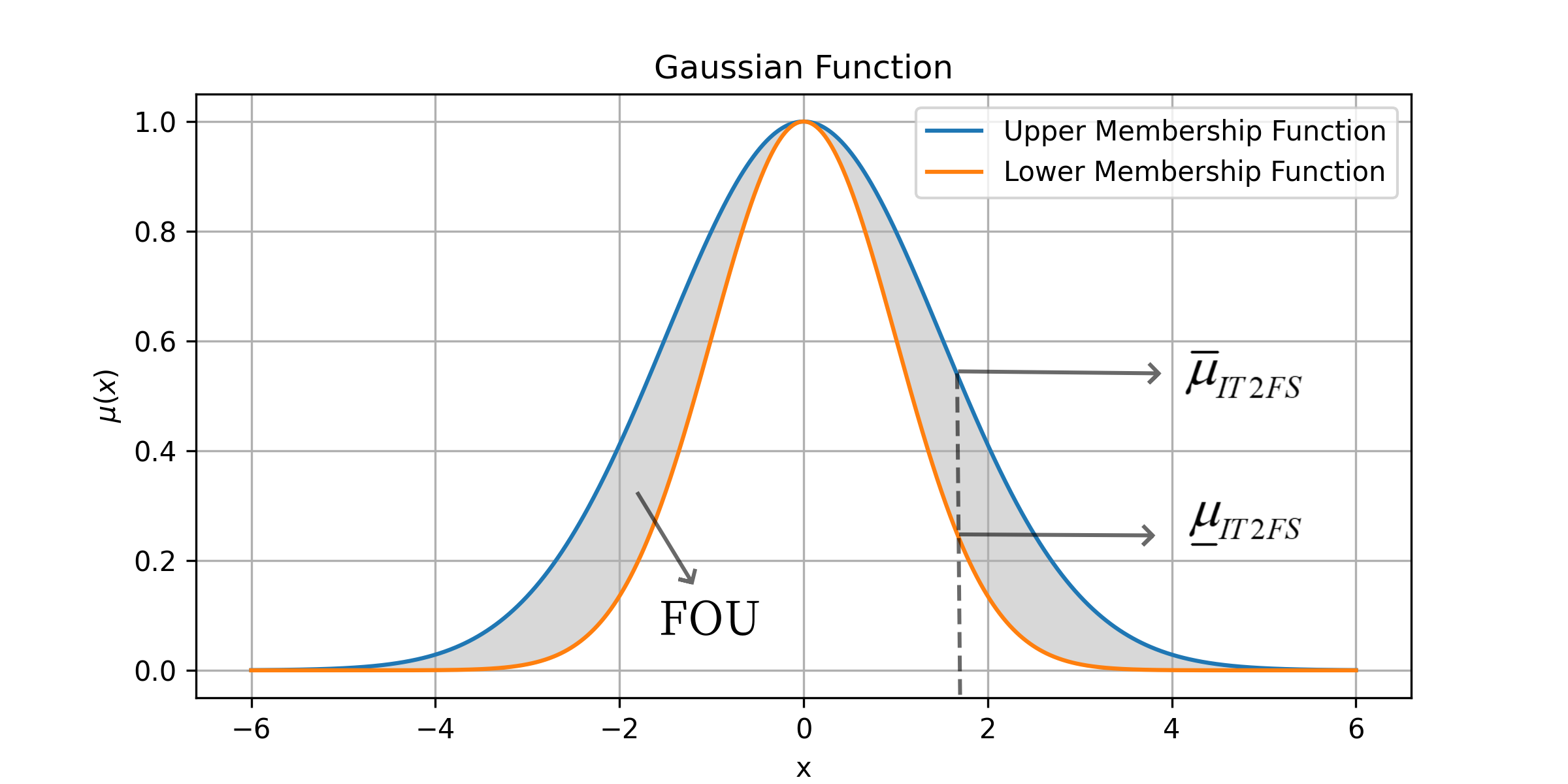} 
	\caption{Standard interval Gaussian membership function.}
	\label{fig:gaussian-membership-function}
\end{figure}

Figure \ref{fig:gaussian-membership-function} illustrates the standard interval Gaussian membership function, where the shaded region is referred to as the Footprint of Uncertainty (FOU).

The interval variance and mean, serving as decisive parameters that determine the membership degree of fuzzy sets, play a pivotal role in the overall performance of the model. To address the limitation of traditional static membership functions in adequately capturing dynamic variations, this study introduces a real-time dynamic adjustment mechanism for parameters in the IT2FIS-GARCH model. For the time-varying parameter IT2FIS, the proper dynamic adjustment of interval variance and mean during both training and application phases constitutes the core procedure to ensure the model meets practical requirements. The estimation process for interval variance is detailed in Subsubsection \ref{Interval estimation of variance}, while the methodology for mean estimation is thoroughly elaborated in Subsubsection \ref{Estimation of the mean}.

\subsubsection{Estimation of the mean}
\label{Estimation of the mean}
The Gaussian membership function, as a mathematical formulation based on normal distribution, has another crucial parameter: the mean. The selection of the mean significantly influences both the function's overall shape and its performance. When the mean is properly set within the normal distribution range of the data, the Gaussian membership function can more effectively focus on normal data points. This section will elaborate in detail on the mean determination process, beginning with an introduction to the fundamental concept of the $3\sigma$ criterion.

In a normal distribution, approximately $68\%$ of the data falls within ±1 standard deviation of the mean, i.e., $[C - \sigma, C + \sigma]$; Approximately $95\%$ of the data falls within ±2 standard deviations of the mean, i.e., $[C - 2\sigma, C + 2\sigma]$; Approximately $99.7\%$ of the data falls within ±3 standard deviations of the mean, i.e., $[C - 3\sigma, C + 3\sigma]$; Based on this rule, any data point exceeding $±3\sigma$ from the mean is generally considered an outlier or extreme value.

According to the $3\sigma$ rule, a reasonable data range can be determined, i.e., $[C - 3\sigma, C + 3\sigma]$. Within this range, the vast majority of data points are included, ensuring that the Gaussian membership function accurately reflects the normal distribution characteristics of the data. Based on this, this paper uses the $3\sigma$ rule to determine the mean parameter of the Gaussian membership function.

In constructing the prediction model described in this study, we have determined the mean based on historical datasets. This mean value can be directly utilized as a parameter during the model training phase. However, when the model enters the prediction phase to handle yet-to-be-generated future data, the mean becomes an unknown quantity. To ensure prediction accuracy, we must estimate the data mean for future time points.

During the fuzzy inference system's prediction process, dynamically setting the current time point's mean as the system's predicted value from the previous moment serves as an effective method to adapt the membership function for time series forecasting needs. As the system operates and predictions progress, the membership function automatically adjusts its center point to better reflect the data characteristics at the current moment. Therefore, incorporating the estimation of future data means during the design and establishment of the prediction model is essential to maintain the model's comprehensive predictive capability during the forecasting phase.

\subsubsection{Establishment of the rule}
\label{Establishment of the rule}
The rule base of the IT2FIS-GARCH model proposed in this paper consists of fuzzy rules in the following form:
\begin{align}
	&\text{If } x_{1} \text{ is } \tilde{A}_{x_{1}}^{i} \text{ and } x_{2} \text{ is } \tilde{A}_{x_{2}}^{i} \text{ and } \ldots \text{ and } x_{t} \text{ is } \tilde{A}_{x_{t}}^{i} \text{ then } \notag \\
	&Y^{i} = a_{0}^{i} + a_{1}^{i}x_{1} + a_{2}^{i} x_{2} + \ldots + a_{t}^{i}x_{t}= a_{0}^{i} + \sum_{j=1}^{t} a_{j}^{i} x_{j}
\end{align}
Where $\tilde{A}_{x_{j}}^{i}$ is the fuzzy antecedent of the input data $x_{j}(j=1,...,t)$ for the $i$-th rule; $Y^{i}$ is the output result of the $i$-th rule; $a_{0}$ , $a_{j}^{i}(j=1,...,t)$ are the consequent parameters of the $i$-th rule. The activation strength $F^{i}$ of the $i$-th rule is calculated using the product method, as shown in the following formula:
\begin{equation}
	F^{i} = \mu_{\tilde{A}{x_1}}^{i} \textperiodcentered \mu_{\tilde{A}{x_2}}^{i} \textperiodcentered...\textperiodcentered \mu_{\tilde{A}_{x_t}}^{i}
	\label{eq:F}
\end{equation}
Here, "\textperiodcentered" denotes multiplication, and $\mu_{\tilde{A}_{x_j}}^{i}$ represents the degree to which the input data $x_j$ in rule $i $ belongs to a set $\tilde{A}_{x_j}(j=1,2,...,t)$, that is, the product of the output results of the fuzzy layer.

Assuming the sliding window size is $W$, the input data sequence at time $t$ is $[x_{t-W+1}, \ldots, x_{t-1}, x_t]$. To accurately describe the degree of membership of each data point, it is divided into $M$ sets. It is important to note that the number of rules that can be activated by the input data at time $t$ in the rule base is $M^{W}$.

During the training of IT2FIS, rule selection and optimization constitute critical steps. To enhance system performance and accuracy, this paper adopts a retention strategy that filters rules based on their firing strengths. Specifically, while the training phase generates numerous fuzzy rules, directly employing all created rules may introduce redundancy or unnecessary information. To address this issue, we retain only the rule with the maximum firing strength ($F_{max}$) at each time step. The firing strength, which indicates how well a rule matches the given inputs, serves as a key metric for evaluating rule importance. By exclusively selecting rules with maximum firing strengths, we construct a concise yet highly efficient rule base. Thus, firing strength-based rule filtering proves to be an effective method for optimizing the rule base during IT2FIS training.

\section{The Prediction Process of the IT2FIS-GARCH Model}
\label{IT2FIS-GARCH model}
This article considers a system with $m $ inputs $ X = \left[ x _ { 1 } , x _ { 2 } , \ldots , x _ { m } \right]$ and $n $outputs $Y =\left[ y _ { 1 } , y _ { 2 } , \ldots , y _ { n } \right] $.

\begin{figure}[H] 
	\centering 
	\includegraphics[width=0.9\textwidth]{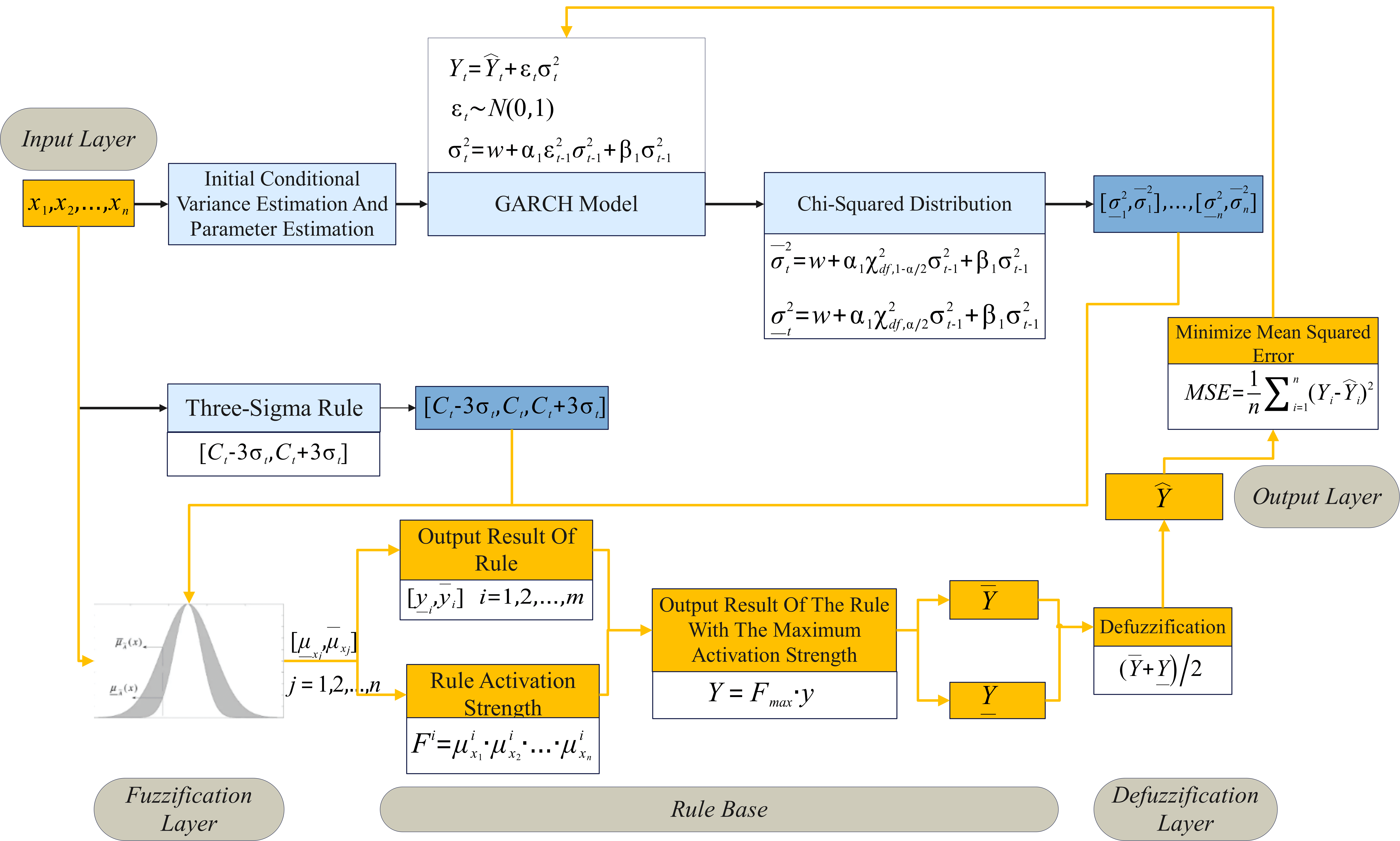} 
	\caption{The IT2FIS-GARCH model. For a dataset with $n$ data points, the variance interval estimate $[\underline{\sigma}_{t}^{2},\overline{\sigma}_{t}^{2}] (t = 1,\cdots,n)$ is first obtained according to the chi-square distribution theory. At the same time, the mean value $C_t$ of the dataset is estimated using the 3$\sigma$ rule. Then, the membership degree intervals of the input data are obtained using the Gaussian membership function. A rule base is used to match the input data to rules, and the interval values of the fuzzy output are then transformed into specific values $\hat{Y}$. During the training process, the predicted value at the final layer is minimized against the true value using the mean squared error operation, thereby training the parameters of the IT2FIS-GARCH model. In the prediction process, the predicted value $\hat{Y}$ from the last layer of the IT2FIS-GARCH model is output.} 
	
	\label{fig:IT2FIS-GARCH算法} 
\end{figure}

Figure \ref{fig:IT2FIS-GARCH算法} shows in detail the specific content and implementation steps of the IT2FIS-GARCH model, including key steps such as model construction, parameter optimization, and output of the final prediction results.

\subsection{Input layer}
\label{Input Layer}
As shown in Figure \ref{fig:IT2FIS-GARCH算法}, the first layer of the IT2FIS-GARCH model is designed as the input layer. The dataset $\tilde { X }$ is divided into two subsets: the training set $\tilde { X }_{train}$ and the testing set $\tilde { X }_{text}$. The training set is used to fit the IT2FIS-GARCH model and adjust the parameters, while the testing set is used to evaluate its prediction performance. Suppose the sliding window size is $W$, meaning that the current value is related to the previous $W$ time steps. The prediction result of the IT2FIS-GARCH model at time $t$ is as follows:
\begin{equation}
	\hat{Y} = model(x_{t-W+1}, \ldots, x_{t-1}, x_t) = \hat{x}_{t+1}
\end{equation}
Here, $ model $ represents the IT2FIS-GARCH prediction model, which takes the data points within the sliding window as input and outputs the predicted value $ \hat{Y} $ for the next time step.
\subsection{Fuzzification layer}
\label{Fuzzification Layer}
As shown in Figure \ref {fig:IT2FIS-GARCH算法}, the second layer of the IT2FIS-GARCH model is the fuzzification layer. At this level, the membership degree of the input data is represented in interval form. Input the preprocessed data into the second layer for fuzzification, and output it as the interval membership value of the input data. Using Gaussian membership function to calculate the interval membership value of the input data at time $t $, the calculation formula is as follows:
\begin{equation}
	\overline { \mu } _ { t } ( x_{t} , \overline { \sigma }_{t} , c_{t} ) = e ^ { - \frac { ( x_{t} - c_{t} ) ^ { 2 } } { 2 \overline { \sigma } ^ { 2 }_{t} } }\label{eq:equation9}
\end{equation}
\begin{equation}
	\underline{\mu} _ { t } ( x_{t} , \underline{ \sigma_{t} } , c_{t} ) = e ^ { - \frac { ( x_{t} - c_{t} ) ^ { 2 } } { 2 \underline { \sigma } ^ { 2 }_{t} } }\label{eq:equation10}
\end{equation}
Among them, $c_ {t} $, $\overline{\sigma} _ {t} $, $\underline{\sigma} _ {t} $ are respectively the center and upper and lower standard deviations of the Gaussian membership function of the input data $x_ {t} $ at time $t $.

In heteroscedastic time series forecasting, to obtain the upper and lower membership values at time $t+1 $ (i.e., [$\underline{\mu}_{t+1}$, $\overline{\mu}_{t+1}$]), the variance at that time point needs to be estimated within an interval. Suppose the sliding window size is $W$, and take the input data $ X_{i} = [x_{i-W+1},\ldots,x_{i-1},x_{i}]$ at time $t$ as an example. First, assume a confidence level of $ 95\% $. Then, the chi-square quantile is determined based on the chi-square quantile table, as given by equation \eqref{eq:chi}. Finally, based on formula \eqref{eq:qujian}, the upper and lower standard deviations at time $t+1$, denoted as $ [\underline{\sigma}_{t+1}, \overline{\sigma}_{t+1}]$, are obtained. According to the $3\sigma$ rule, the mean of the input data at time $t+1$ is given by $[\tilde{C}_{t+1} - 3\sigma_{t+1}, \tilde{C}_{t+1}, \tilde{C}_{t+1} + 3\sigma_{t+1}]$. The parameter identification process for formula \eqref{eq:tiaojianyifangchafangcheng} is as shown in Subsection \ref{Estimation of GARCH model parameters}.

After obtaining the mean and the upper and lower standard deviation values of the input data, the interval membership values $ [\underline {\mu} _ { t+1 },\overline {\mu} _ { t+1 }]$ for the input data at time $t+1$ are obtained using formulas \eqref{eq:equation9} and \eqref{eq:equation10}. This interval membership is used as one of the inputs to the third layer of the IT2FIS-GARCH model.

\subsection{Rule base}
\label{Rule Base}
As shown in Subsubsection \ref{Establishment of the rule}, this article adopts a retention strategy, which only retains the rule with the maximum activation strength ($F_{max}$) at each moment. The final fuzzy output result of the rule layer can be obtained through the following formula:
\begin{equation}
	\begin{aligned}
		\overline{Y}^{pred} = Y^{l}\overline{F}_{max}\\
		\underline{Y}^{pred} = Y^{r}\underline{F}_{max}
	\end{aligned}
	\label{eq:output}
\end{equation}
In this context, $\overline{Y}^{pred}, \underline{Y}^{pred}$ represent the upper and lower bounds of the fuzzy output, respectively. When calculating the upper bound $\overline{Y}^{pred}$, the output value $Y^{l}$ corresponding to the rule with the maximum activation strength (i.e., the $l$-th rule) is used. Similarly, when calculating the lower bound $\underline{Y}^{pred}$, the output value $Y^{r}$ corresponding to the rule with the maximum activation strength (i.e., the $r$-th rule) is used.

Taking the upper bound of the interval as an example, the activation strengths of all the rules activated by the input data at time $t$ are first calculated according to formula \eqref{eq:F}. Then, the rule with the maximum activation strength ($\overline{F}_{max}$) is selected from these rules. Finally, using the selected rule and its corresponding output (formula \eqref{eq:output}), the upper bound of the fuzzy output $\overline{Y}^{pred}$ can be obtained. Similarly, the lower bound of the fuzzy output $\underline{Y}^{pred}$ can be obtained, resulting in a complete fuzzy output interval $[\underline{Y}^{pred},\overline{Y}^{pred}]$. This method not only improves the prediction accuracy but also enhances the model’s ability to handle uncertainty and fuzziness. It allows the IT2FIS-GARCH model to accurately predict GARCH-type time series data and output the corresponding prediction results. The fuzzy output from the third layer is then used as input for the fourth layer.

\subsection{Defuzzification layer}
After matching the rules in the rule library, the predicted output interval of our model is finally obtained, which consists of two key boundaries, $ [\underline{Y}^{pred},\overline{Y}^{pred}] $. Among them, $\overline {Y} ^ {pred} $ and $\underline {Y} ^ {pred} $ are the upper and lower bounds of the fuzzy results when the model is used for prediction in this paper. In order to express the prediction results more intuitively, this article adopts the interval midpoint method, a deblurring technique, to convert the fuzzy results into clear numerical values. The formula is as follows:
\begin{equation}
	Y^{final} = \dfrac{\overline{Y}^{pred}+\underline{Y}^{pred}}{2}
	\label{eq:qujianzhongdian}
\end{equation}

Use the formula \eqref {eq:qujianzhongdian} to convert the interval type output into a single, specific prediction value, which is used as the expression form of the final prediction result and as the input of the model output layer.
\subsection{Output layer}
The IT2FIS-GARCH model implements the final output $Y ^ {final} $ at the output layer. The model constructed in this article adopts a rolling prediction strategy to update input data and achieve multi-step prediction $ (Y^{final}=[x_{t+1},x_{t+2}...,x_{t+steps}])$, where $steps $ represents the prediction step size set by the model in multi-step prediction.

In the one-step prediction framework, the model uses the window data at time $t $ to predict the data values at time $t+1 $. This prediction process first iteratively calculates the variance prediction interval value at time $t+1 $ by using formulas \eqref {eq:chi} and \eqref {eq:qujian} on the window data. Subsequently, the $3 \sigma $ criterion is applied to determine the mean at time $t+1 $. Next, use a variable parameter Gaussian membership function to calculate the interval membership values. The IT2FIS-GARCH model then uses these interval membership values and input window data to calculate the predicted value at time $t+1 $ through its internal logic and rules (as shown in Section \ref{IT2FIS-GARCH model}).

When performing multi-step prediction in IT2FIS-GARCH, the above process becomes more complex but orderly. Firstly, the predicted values of the IT2FIS-GARCH model are used as the new mean in the GARCH model mean equation to reflect the dynamic changes in time series data. At the same time, the predicted value at that moment is also considered as a new data point added to the sliding window, updating the data window while keeping the window size constant. The updated window data is used to recalculate the interval variance value (i.e. Subsubsection \ref{Interval estimation of variance}), mean estimation (i.e. Subsubsection \ref{Estimation of the mean}), membership value calculation (i.e. Subsection \ref{Fuzzification Layer}), and other processes for the next time point, ultimately achieving prediction for the next time point. This rolling prediction mechanism ensures that each step of the prediction is based on the latest available information, thereby improving the accuracy and timeliness of the prediction. In the process of continuous iteration, the IT2FIS-GARCH model can continuously predict the values of multiple future time points, achieving multi-step rolling prediction. Assuming the window size is $W $, the performance of the IT2FIS-GARCH model for multi-step rolling prediction is as follows:
\begin{equation}
	\begin{aligned}
		\hat{Y}_{t+1} &= model(x_{t-W+1}, \ldots, x_{t-1}, x_t) = \hat{x}_{t+1},\\
		\hat{Y}_{t+2} &= model(x_{t-W+2}, \ldots, x_{t}, \hat{x}_{t+1}) = \hat{x}_{t+2},\\
		\cdots,\\
		\hat{Y}_{t+steps} &= model(\hat{x}_{t+stpes-W}, \ldots, \hat{x}_{t+steps-2}, \hat{x}_{t+steps-1}) = \hat{x}_{t+steps}.
	\end{aligned}
\end{equation}
Among them, $steps $ represents the prediction step set by the IT2FIS-GARCH model during the prediction process. $\{\hat{Y}_{t+1},\hat{Y}_{t+2},\cdots,\hat{Y}_{t+steps}\}$ represents the output of the IT2FIS-GARCH model during multi-step prediction.

In order to improve the performance of the IT2FIS-GARCH model, this paper adjusts and optimizes the various parameters of the model using the least squares method. The detailed code for the application phase of the IT2FIS-GARCH model is provided in Algorithm 4.1.

\begin{table}[h]
	\centering
	\footnotesize
	\begin{tabular}{l}
		\hline
		\textbf{Algorithm 4.1} IT2FIS-GARCH:The code for the IT2FIS model testing phase. \\
		\hline
		\textbf{Input:}New data $\tilde { X }_{test} = \{x_1, x_2,\ldots, x_i\}$, sliding window size $W $, predicted steps $steps $.\\
		\textbf{Output:}Out of sample predicted values.\\
		1.Replace the outliers in $\tilde {X} _ {test} $ using the linear difference method;\\
		
		2.Obtain the optimal parameters ($w^{best},\alpha^{best},\beta^{best},a_0^{best},a_1^{best},\ldots,a_t^{best}$) of the model\\
		  using Algorithm 3.1;\\
		3.A random number $\varepsilon $ with a length of $len (\tilde {X} _ {test}) $ that follows a normal distribution;\\
		
		\textbf{for} $t=1$ \textbf{to} $len(\tilde { X }_{test})$\\
		\hspace{2em} Add the current value $x_ {t} $ to the window $win $;\\
		\hspace{2em} \textbf{if} $len(win)\textgreater W$ \textbf{then}\\
		\hspace{4em} Delete the oldest data in the window;\\
		\hspace{2em} \textbf{end if}\\
		\hspace{2em} Obtain the interval value of $\varepsilon ^ 2 $ based on the chi square distribution table;\\
		\hspace{2em} Predict the variance value at time $t $ by $\sigma_{t}^2 = w^{best} +  \alpha^{best} \varepsilon_{t-1}^2\sigma_{t-1}^2 +  \beta^{best} \sigma_{t-1}^2$;\\
		\hspace{2em} Obtain the interval value of variance [$\underline{\sigma}_{t}^2$, $\overline{\sigma}_{t}^2$];\\
		\hspace{2em} Update the mean of the predicted values based on the $3\sigma$ criterion\\
		\hspace{2em} and the IT2FIS-GARCH model;\\
		\hspace{2em} Calculate the interval membership value of data based on\\
		\hspace{2em} Gaussian membership function $[\underline{\mu} _ { t },\overline{\mu} _ { t }]$;\\
		\hspace{2em} Match fuzzy rules and output the final predicted value $\hat{Y}_t = a_{0}^{best} + \sum_{j=1}^{t} a_{j}^{best} x_{j}$;\\
		\hspace{2em} \textbf{for} 1 \textbf{to} $steps$\\
		\hspace{4em} Update window data, delete the oldest data, add the latest predicted values;\\
		\hspace{4em} Recalculate the variance interval value, mean,\\
		\hspace{4em} and interval membership value of the window data;\\
		\hspace{4em} Output the predicted value of future steps $\hat{Y}_ {t+steps}$,\\
		\hspace{4em} store the predicted values in the predicted value list $\hat{Y} ^ {final} $;\\
		\hspace{2em} \textbf{end for}\\
		\textbf{end for}\\
		\textbf{return} $\hat{Y}^{final}$\\
		
		\hline
	\end{tabular}	
\end{table}

\section{Experiment and Result Analysis}
\label{experiment and result analysis}
To validate the performance of the IT2FIS-GARCH model proposed in this article, experiments were conducted on ten datasets based on five evaluation metrics. The dataset used in the comparative experiment was sourced from the UCI database and the Data.gov data website. The dataset is "AirQualityUCI", "Average\_Daily\_Traffic\_Counts", "Data", "Death", "diabetesl\_2016", "energydata\_complete", "ex\_20", "household\_power\_cons-\\umption", "PRSA\_Data\_Dongsi", "Traffic". The data supporting this study are openly available in the UCI Machine Learning Repository and Data.gov. A complete list of datasets, including their access links and references, is provided in [Data Availability Statement]. Compare the generalized autoregressive conditional heteroskedasticity Takagi Sugeno Kang (GARCH-TSK) model, fixed variance time series model (Fixed Variance IT2FIS) \cite{khairuddin_generating_2022}, generalized autoregressive conditional heteroskedasticity gated recurrent unit (GARCH-GRU) model \cite{ali_traffic_2023}, and long short-term memory neural network (LSTM-IT2FIS) \cite{petrozziello_deep_2022}. In order to effectively describe the performance and accuracy of the IT2FIS-GARCH model in predicting GARCH-type time series, the following evaluation metrics are used in this section.

Mean Square Error (MSE): MSE is used to measure the square of the average difference between predicted values and actual observed values, calculated using the formula \eqref{eq:mse}.
The smaller the MSE value, the smaller the difference between the predicted results of the model and the actual observed values, indicating a stronger predictive ability of the model.

Root Mean Square Error (RMSE): RMSE is the arithmetic square root of MSE, reflecting the degree of deviation between the true value and the predicted value. The calculation formula is as follows:
\begin{equation}
	\text{RMSE} = \sqrt{\text{MSE}} = \sqrt{\frac{1}{n} \sum_{i=1}^{n} (y_i - \hat{y}_i)^2}
\end{equation}

Mean Absolute Error (MAE): MAE is one of the commonly used indicators to evaluate the performance of prediction models, used to measure the average absolute error between predicted values and actual observed values. It is the average of the absolute values of prediction errors, and for a dataset containing $n $ samples, its calculation formula is:
\begin{equation}
	\text{MAE} = \frac{1}{n} \sum_{i=1}^{n} |y_i - \hat{y}_i|
\end{equation}
Where $y_i $ is the actual observation value of the $i$-th sample $ \hat {y}_i $ is the predicted value of the $i$-th sample.

Mean Absolute Percentage Error (MAPE): It is a commonly used indicator in regression analysis to evaluate the accuracy of predictive models. It is used to measure the difference between predicted and actual values, expressed in percentage form. MAPE calculates the average absolute percentage error between predicted and actual values. Its mathematical formula is:
\begin{equation}
	MAPE = \frac{1}{n} \sum_{i=1}^{n} \left| \frac{A_i - F_i}{A_i} \right| \times 100
\end{equation}
Among them, $A_i $ is the $i $th actual observation value $ F_i $ is the $i $th predicted value $ n $ is the total number of samples.

Determination coefficient ($R ^ 2 $): $R ^ 2 $ is a statistical indicator used to evaluate the goodness of fit of a regression model $R^2$ measures the proportion of dependent variable variation that the regression model can explain. The value of  $R^2$ is between 0 and 1. The closer it is to 1, the better the fitting effect of the model, and the closer it is to 0, the worse the fitting effect of the model. Its mathematical formula is:
\begin{equation}
	R^2 = 1 - \frac{SS_{res}}{SS_{tot}}
\end{equation}
Among them, $SS_{res}$ is the sum of squared residuals, representing the sum of squared differences between the predicted and actual values of the model $ SS_{tot} $ is the total sum of squares, representing the sum of squares of the difference between the actual value and the mean of the actual value.

In the implementation process of the $n $ step prediction in the IT2FIS-GARCH model, in order to more intuitively display the prediction error results, this paper uses the mean of the prediction error values for future $n $ steps at each moment as the representative error value for that moment.

\subsection{Data preprocessing and model analysis}
\label{Data preprocessing and model analysis}
As this article focuses on the prediction problem of heteroscedastic time series, linear interpolation method is chosen to handle outliers or "NA" values. The formula is as follows:
\begin{equation}
	x _ { t } = \frac { \sum _ { i = 1 } ^ { n } ( x _ { t - i } + x _ { t + i } ) } { 2 n }
	\label{eq:1}
\end{equation}
Among them, $x_ {t}$ represents the data value at the current time $t $, and the average of $n $ time values on the left and right sides is taken as the replacement value for the current time.

This article also uses a statistical method for analyzing time series: Ljung Box test\cite{ljung-measure_1978}, which aims to evaluate whether the model can comprehensively capture and reflect the dynamic features in the data. The formula is as follows:
\begin{equation}
	Q = n ( n + 2 ) \sum _ { r = 1 } ^ { h } ( n - r ) ^ { - 1 } \rho ^ { 2 } ( r )
	\label{eq:3}
\end{equation}
Where $n $ is the input sample size, $ h $ is the set lag period, used to specify the lag order to be considered.

Hypothesis of Ljung Box test: null hypothesis (H0): The autocorrelation coefficient of the residual sequence of the time series at a lag period of $h $ is zero, i.e. ($\rho(r)=0 $), where $\rho(r) $ is the autocorrelation coefficient of the sample data at different lag periods.
Alternative hypothesis (H1): The residual sequence of a time series exhibits autocorrelation at least for some lag periods ($h $), i.e. ($\rho(r) \neq 0 $). If the Ljung Box test results show significant autocorrelation in the residuals, it indicates that the model has not fully captured dynamic features. Therefore, the application of this method is particularly important in time series prediction problems.

\subsection{Comparison of mean square error experimental results}
\label{MSE}
\begin{figure}[H]
	\centering
	\begin{subfigure}[t]{0.49\textwidth} 
		\centering
		\includegraphics[height=5cm]{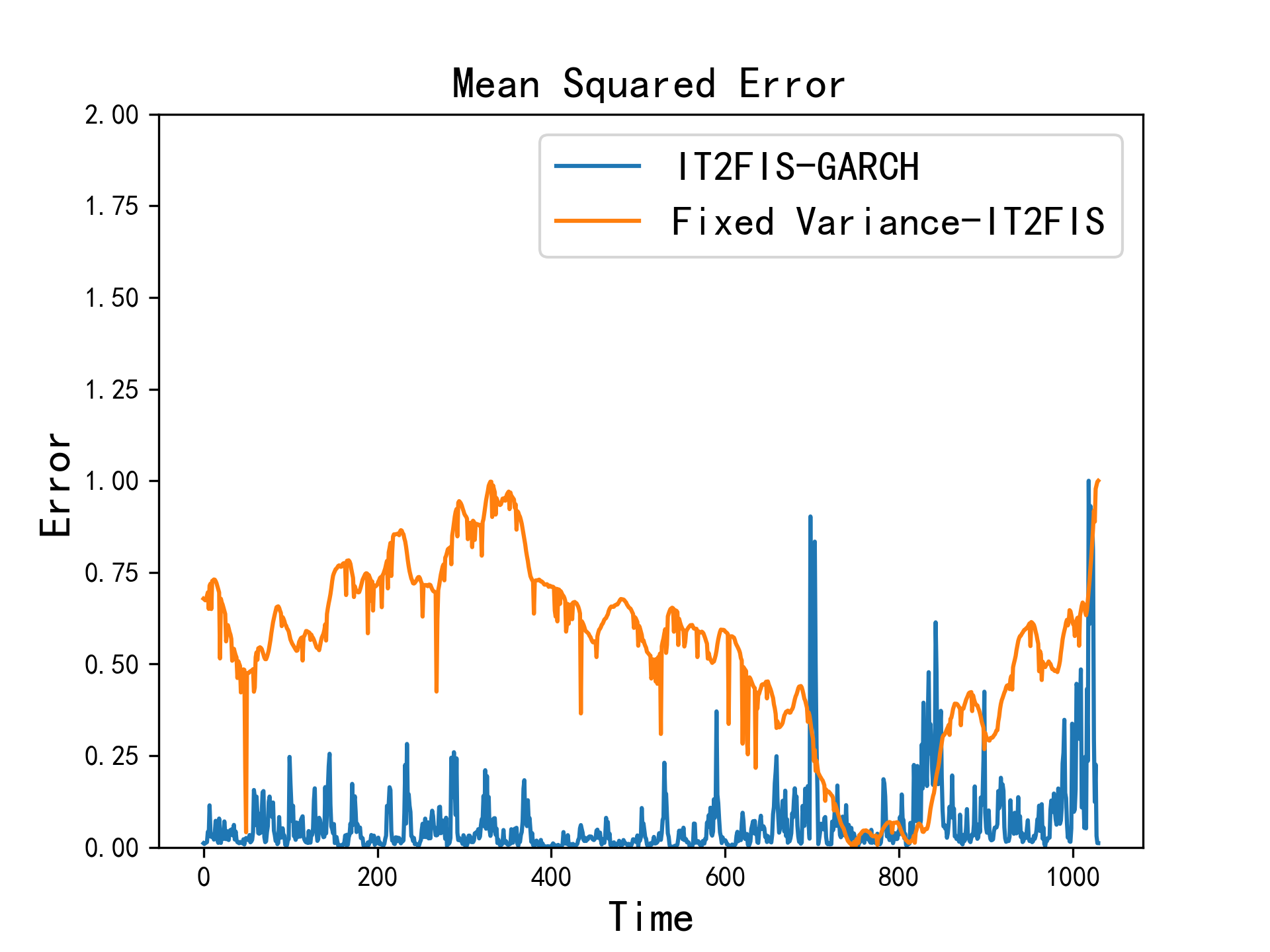}
		\caption{MSE of the Prediction for the 'Data' Dataset}
		\label{fig:data}
	\end{subfigure}
	\hfill
	\begin{subfigure}[t]{0.49\textwidth} 
		\centering
		\includegraphics[height=5cm]{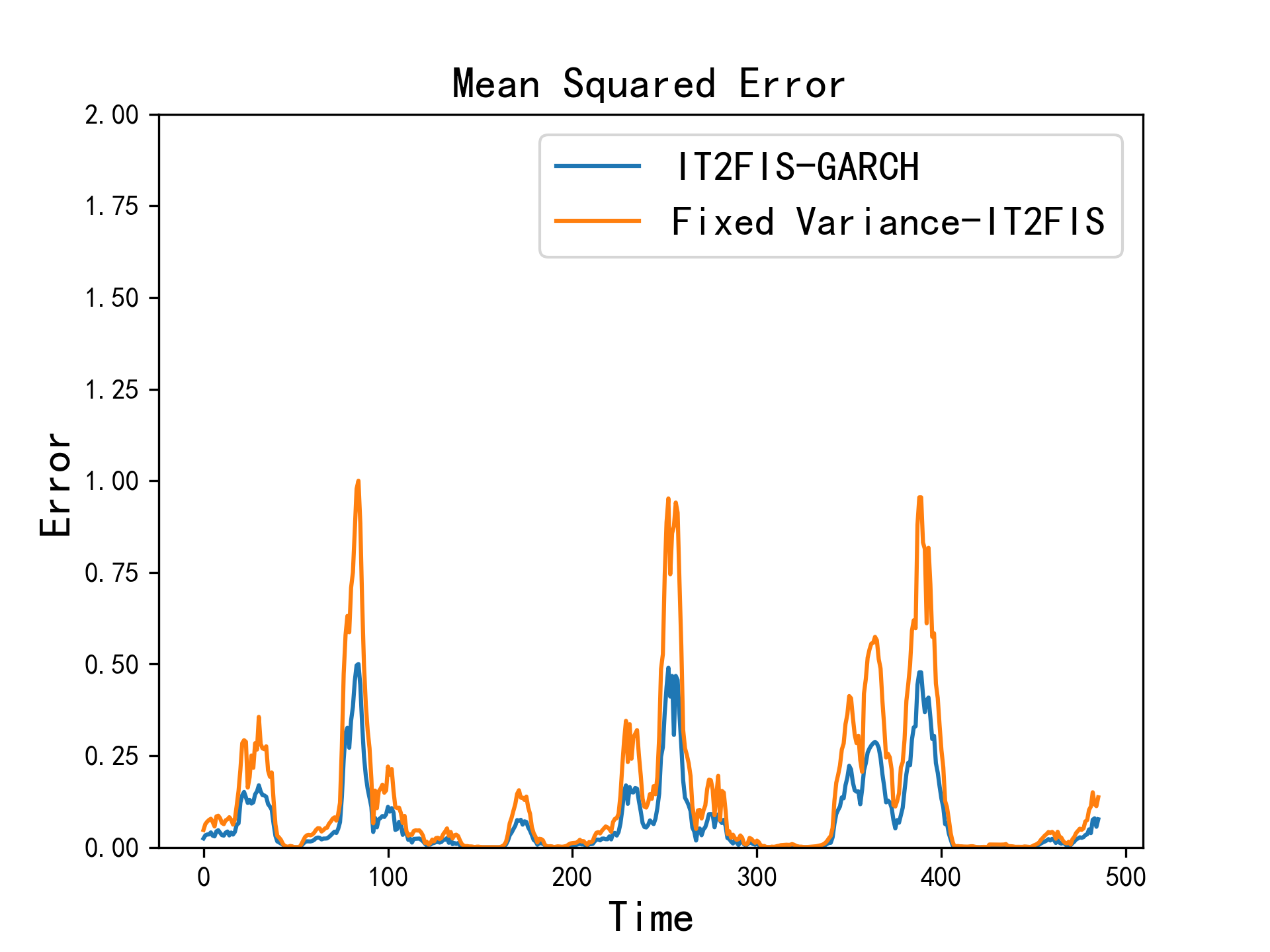}
		\caption{MSE of the Prediction for the 'PRSA\_Data\_Dongsi' Dataset}
		\label{fig:PRSA}
	\end{subfigure}
	\caption{Comparison of IT2FIS-GARCH model and Fixed Variance-IT2FIS model for prediction in different datasets: (a) "Data", (b) "PRSA\_Data\_Dongsi". The blue line represents the IT2FIS-GARCH model, and the orange line represents the Fixed Variance-IT2FIS model.}
	\label{fig:comparison1}
\end{figure}

In this section, 10 datasets are selected and the performance of the proposed model is validated against other models using the existing algorithm (MSE). By observing the predictive results of the model on different GARCH-type time series datasets, comprehensively evaluate and compare its predictive ability. Figure \ref{fig:data} and Figure \ref{fig:PRSA} show the mean square error plots of two time series datasets. Table \ref{tab:MSE} shows the mean square error of different models on ten GARCH-type time series datasets.

The dataset in Figure \ref{fig:data} covers the NSE-NIFTY 50 (National Stock Exchange) index data from January 1, 2008 to December 2, 2021. The dataset in Figure \ref{fig:PRSA} is an hourly air pollutant dataset from the air quality monitoring station of Beijing Environmental Monitoring Center. Evaluate the model proposed in this article against different comparative models based on evaluation indicators.

\renewcommand{\arraystretch}{2}
\begin{table}[H]
	\centering
	\tiny
	\caption{Prediction accuracy of IT2FIS-GARCH model and different comparison models on ten datasets (MSE as evaluation metric)}
	\label{tab:MSE}
	\begin{tabular}{ c c c c c c }
		\hline
		Model & IT2FIS-GARCH & \makecell{Fixed Variance- \\ IT2FIS} & LSTM-IT2FIS & GARCH-GRU & GARCH-TSK \\
		\hline
		
		AirQualityUCI& 0.06163 & \textbf{0.04007} & 0.13238 & 0.10201 & 0.18075 \\
		\hline
		\makecell{Average\_Daily\_ \\ Traffic\_Counts}& \textbf{0.00563} & 0.32703 & 0.32324 & 0.04914 & 0.18656\\
		\hline
		Data& \textbf{0.03732} & 0.53610 & 0.69062 &0.57906& 0.25043 \\
		\hline
		Death& 0.12229 & \textbf{0.02188} & 0.14303 & 0.11582 & 0.11250\\
		\hline
		diabetesl\_2016& \textbf{0.01268} & 0.01711 & 0.32271 & 0.16380 & 0.57061\\
		\hline
		energydata\_complete& \textbf{0.01445} & 0.83804 & 0.44831 & 0.54130 & 0.02836 \\
		\hline
		ex\_20& 0.37552 & 0.55515 & 0.57675 & \textbf{0.27030} & 0.45465 \\
		\hline
		\makecell{household\_power\_ \\ consumption}& \textbf{0.04751} & 0.14992 & 0.17811 & 0.18053 & 0.59342 \\
		\hline
		PRSA\_Data\_Dongsi& 0.07247 & 0.14226 & \textbf{0.02440} &0.06740 & 0.10262 \\
		\hline
		Traffic& \textbf{0.06652} & 0.21001 &0.19972&0.28698& 0.07995\\
		\hline
	\end{tabular}
	
\end{table}

Figure \ref{fig:data} and Figure \ref{fig:PRSA} show the mean square error results of the IT2FIS-GARCH model and Variance-IT2FIS model on the datasets "Data" and "PRSA\_Data\_Dongsi", respectively. By observing Figure \ref{fig:data} and Figure \ref{fig:PRSA}, it can be seen that the error fluctuation range of IT2FIS-GARCH is relatively small. In contrast, fixed variance time series models have poorer predictive performance. Table \ref{tab:MSE} shows the mean square error values for all datasets. From Table \ref{tab:MSE}, it can be seen that the model constructed in this paper has a smaller mean square error than the fixed variance time series model on six datasets. In the prediction task for the dataset "Average\_Daily\_Traffic\_Counts", although the prediction error of the IT2FIS-GARCH model did not reach the minimum value, it was very close to the prediction error obtained by the GARCH-GRU model. The prediction error of interval type-2 fuzzy inference system with fixed variance on the dataset "Death" is smaller. In the prediction task of the dataset "PRSA\_Data\_Don-\\gsi", the ability of long short-term neural networks to capture time series heteroskedasticity is stronger than that of GARCH models.

\subsection{Comparison of root mean square error experimental results}
\label{RMSE}
\begin{figure}[H]
	\centering
	\begin{subfigure}[t]{0.49\textwidth}  
		\centering
		\includegraphics[width=\textwidth]{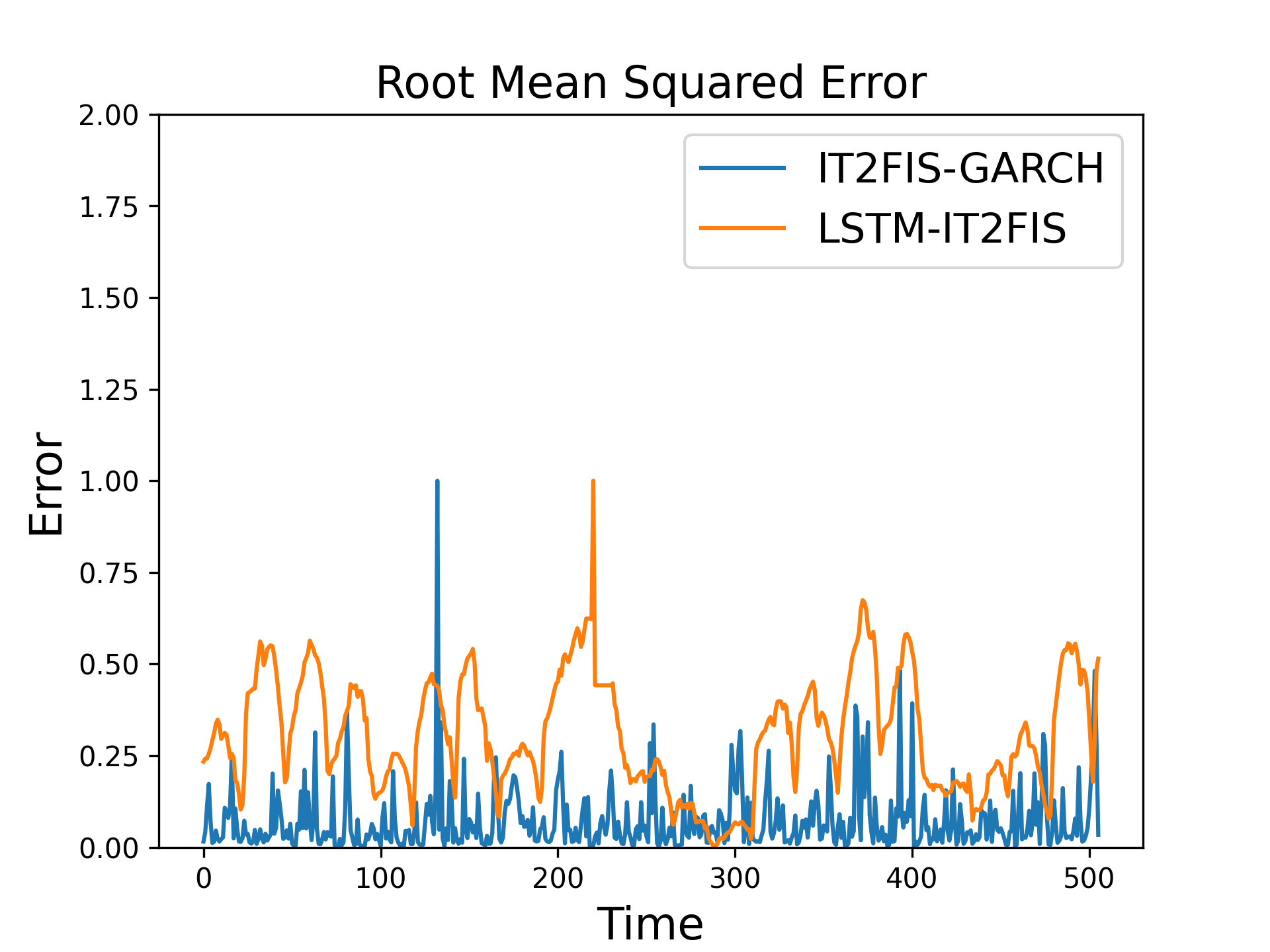}
		\caption{RMSE of the Prediction for the 'AirQualityUCI' Dataset}
		\label{fig:LSTM-air}
	\end{subfigure}
	\hfill
	\begin{subfigure}[t]{0.49\textwidth}
		\centering
		\includegraphics[width=\textwidth]{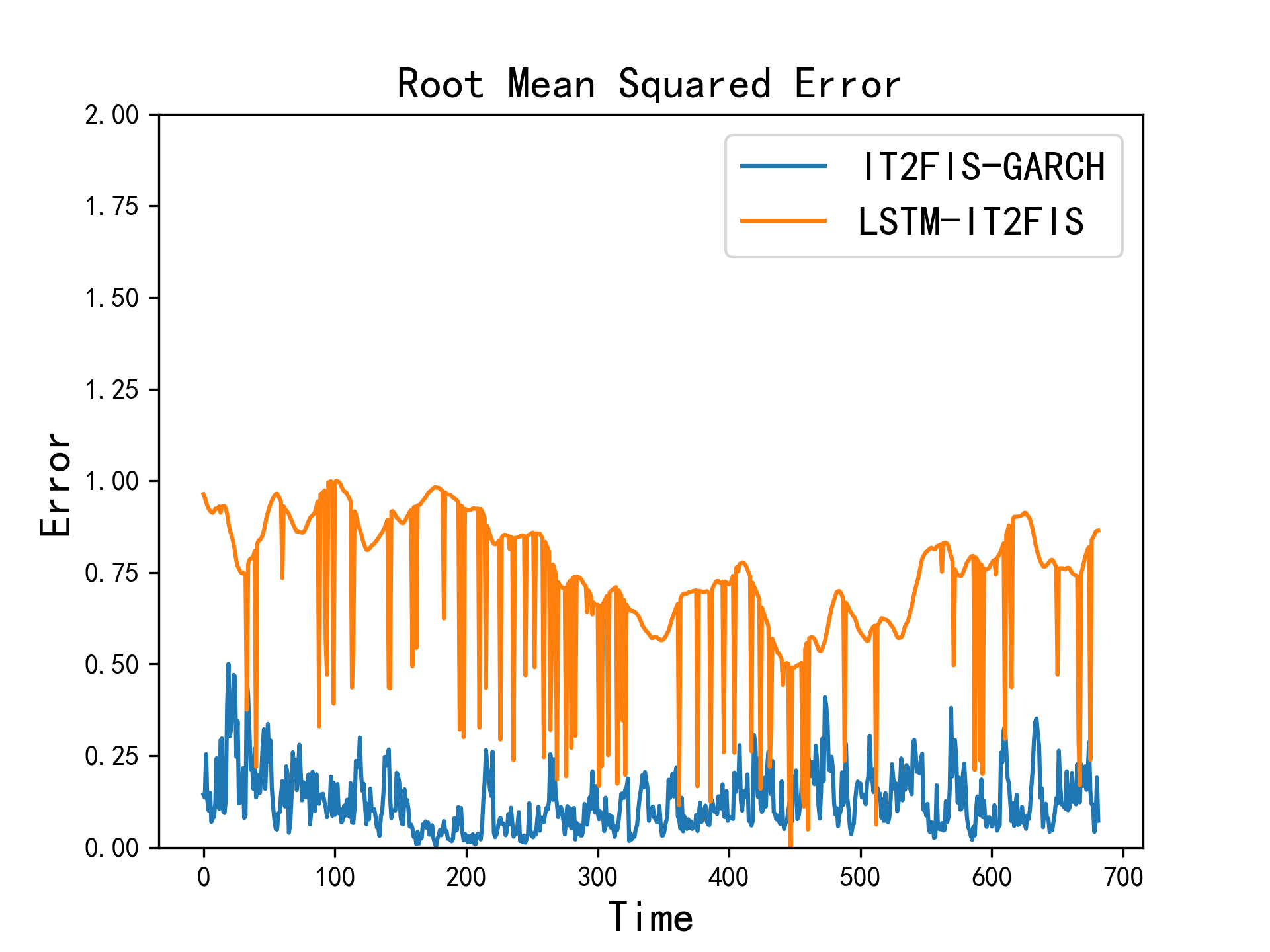}
		\caption{RMSE of the Prediction for the 'Data' Dataset}
		\label{fig:LSTM-data}
	\end{subfigure}
	
	\vspace{1em}  
	\begin{subfigure}[t]{0.6\textwidth}  
		\centering
		\includegraphics[width=\textwidth]{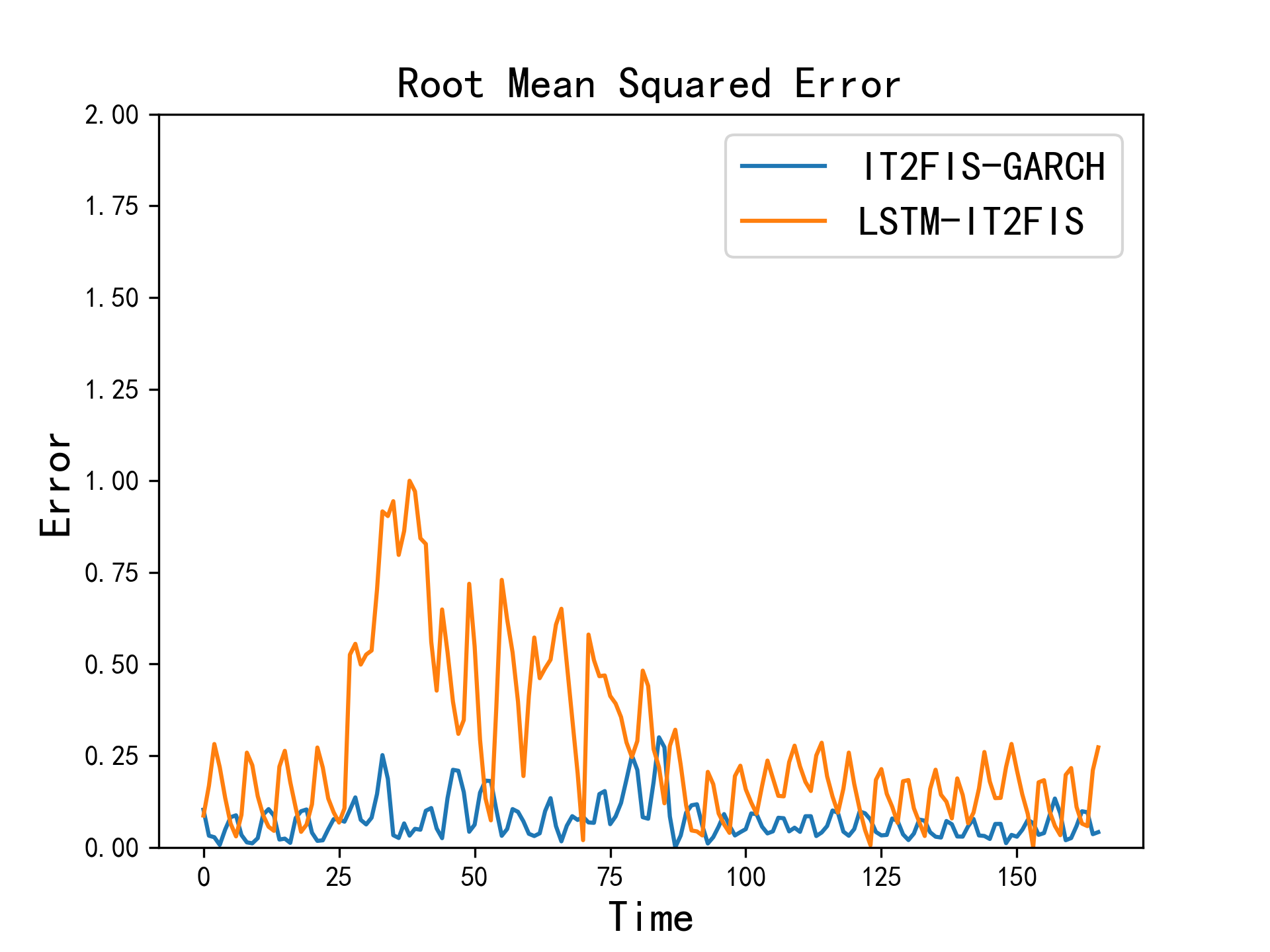}
		\caption{RMSE of the Prediction for the 'household\_power\_consumption' Dataset}
		\label{fig:LSTM-household}
	\end{subfigure}
	
	\caption{Comparison of IT2FIS-GARCH model and LSTM-IT2FIS model for prediction in different datasets: (a) "AirQualityUCI", (b) "Data", and (c) "household\_power\_consumption". The blue line represents the IT2FIS-GARCH model, and the orange line represents the LSTM-IT2FIS model.}
	\label{fig:comparison}
\end{figure}

In this section, the predictive ability of the IT2FIS-GARCH model is compared with the comparison model based on the existing algorithm (RMSE). The dataset in Figure \ref{fig:LSTM-air} is experimental data used to create a low-energy building electrical energy consumption regression model. The dataset in Figure \ref{fig:LSTM-data} uses data from the NSE-NIFTY 50 (National Stock Exchange) index. The dataset of LSTM household in Figure \ref {fig:LSTM-household} records the active energy consumed per minute by electrical devices in the household. This section mainly evaluates the performance of the IT2FIS-GARCH model proposed in this paper in terms of prediction accuracy by comparing the values of root mean square error (RMSE).

\begin{table}[H]
	\centering
	\tiny
	\caption{Prediction accuracy of IT2FIS-GARCH model and different comparison models on ten datasets (RMSE as evaluation metric)}
	\label{tab:RMSE}
	\begin{tabular}{ c c c c c c }
		\hline
		Model & IT2FIS-GARCH & \makecell{Fixed Variance- \\ IT2FIS} & LSTM-IT2FIS & GARCH-GRU & GARCH-TSK \\
		\hline
		AirQualityUCI& \textbf{0.15177} & 0.52918 & 0.30921 & 0.24223 & 0.32348\\
		\hline
		\makecell{Average\_Daily\_ \\ Traffic\_Counts}& \textbf{0.01324} &0.41175&0.40748&0.18383& 0.33106\\
		\hline
		Data& \textbf{0.14270} &0.60933&0.73839&0.65417& 0.38769\\
		\hline
		Death& 0.23945 &\textbf{0.09284}&0.25901& 0.26740& 0.22295\\
		\hline
		diabetesl\_2016& \textbf{0.01897} & 0.02863& 0.41645&0.33559& 0.57697 \\
		\hline
		energydata\_complete& \textbf{0.06144} & 0.90470 &0.53878&0.56847& 0.10112\\
		\hline
		ex\_20& 0.52681 & 0.55716&0.57844&\textbf{0.28956}& 0.45900 \\
		\hline
		\makecell{household\_power\_ \\ consumption}& \textbf{0.03799} & 0.26491& 0.27342&0.35749& 0.66058\\
		\hline
		PRSA\_Data\_Dongsi& 0.14247 & 0.28463&\textbf{0.08033}&0.16755&0.23369 \\
		\hline
		Traffic& \textbf{0.14856}& 0.43354&0.28879 &0.43857& 0.19337\\%
		\hline
	\end{tabular}
	
\end{table}

Taking the IT2FIS-GARCH model and LSTM-IT2FIS model as examples, Figure \ref {fig:LSTM-data}, Figure \ref{fig:LSTM-air}, and Figure \ref{fig:LSTM-household} are the root mean square error results of the two models. It can be observed that the prediction error of the IT2FIS-GARCH model is better than that of the LSTM-IT2FIS model on different time series datasets. This indicates that the IT2FIS-GARCH model has good stability in dealing with data changes and can ensure that the prediction results are within a certain error range. Furthermore, according to the root mean square error values shown in Table \ref {tab:RMSE}, it can be observed that the IT2FIS-GARCH model has a smaller mean error across the seven datasets.

\subsection{Comparison of MAE and MAPE experimental results}
\label{Comparison of MAE and MAPE Experimental Results}

In this section, existing algorithms (MAE, MAPE) are selected to validate the "interval" characteristics of the IT2FIS-GARCH model. Compare IT2FIS-GARCH with different models to verify the ability of "interval" to handle uncertainty. The value of RMSE is affected by significant errors (especially outliers or outliers) in the model prediction process, as the errors are amplified after squaring. And MAE is the average of the absolute values of errors, with the same sensitivity to all errors and not particularly sensitive to large errors. MAPE is presented in percentage form, which can intuitively reflect the size of prediction error and facilitate comparison between different models or datasets. MAPE is based on the proportional error of actual values, so it is more sensitive to real values with smaller values.

Select two different GARCH-type time series datasets, and their MAE results are shown in the following figure. Figure \ref {fig:GRU-Death} shows the suicide mortality rate in the United States by gender, race, Hispanic ethnicity, and age. Figure \ref {fig:GRU-traffic} shows the hourly westbound traffic volume of Interstate 94 at DoT ATR Station 301 in Minnesota.

\begin{figure}[H]
	\centering
	\begin{subfigure}[t]{0.49\textwidth} 
		\centering
		\includegraphics[height=5cm]{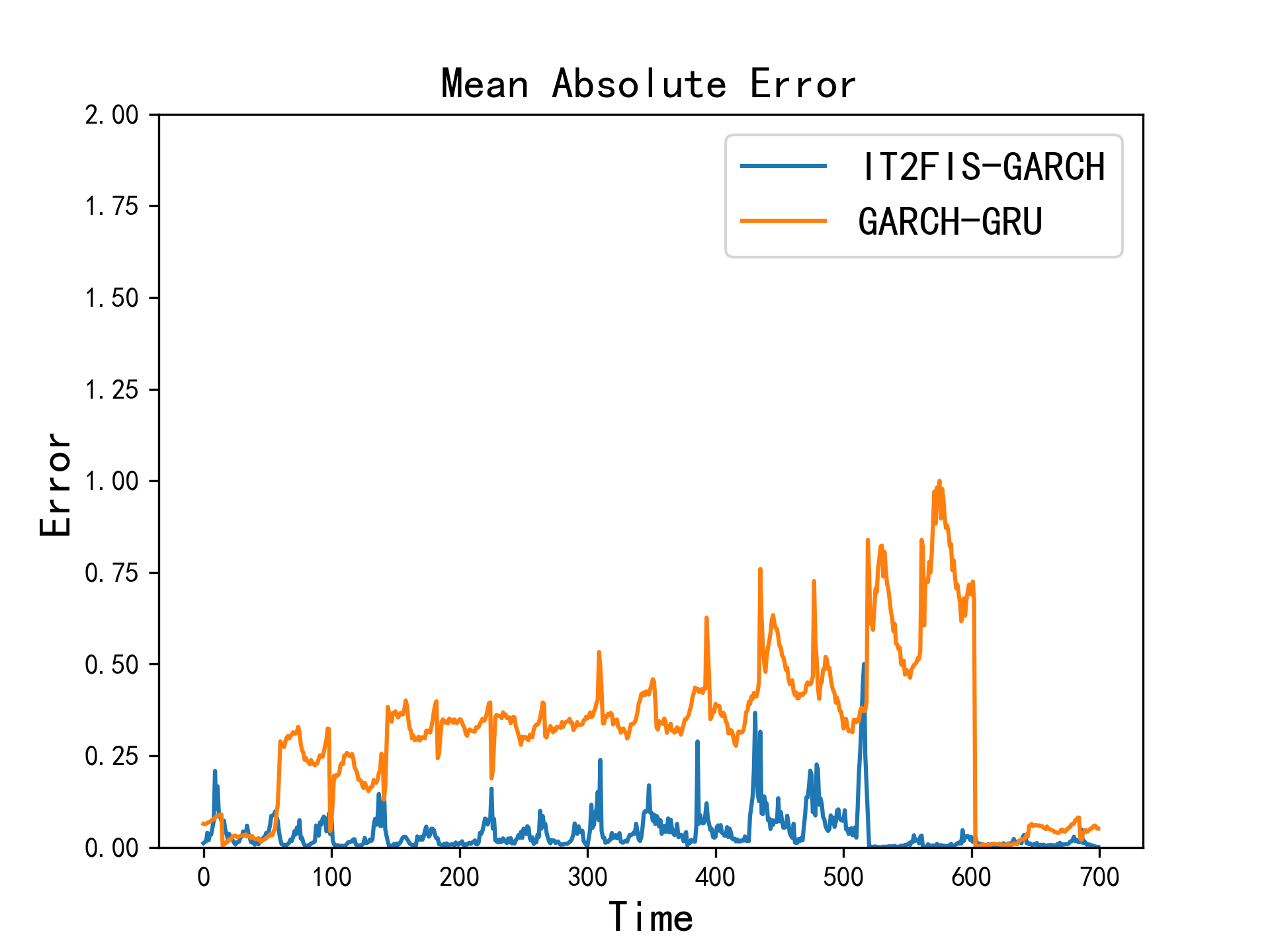}
		\caption{MAE of the Prediction for the 'Death' Dataset}
		\label{fig:GRU-Death}
	\end{subfigure}
	\hfill
	\begin{subfigure}[t]{0.49\textwidth} 
		\centering
		\includegraphics[height=5cm]{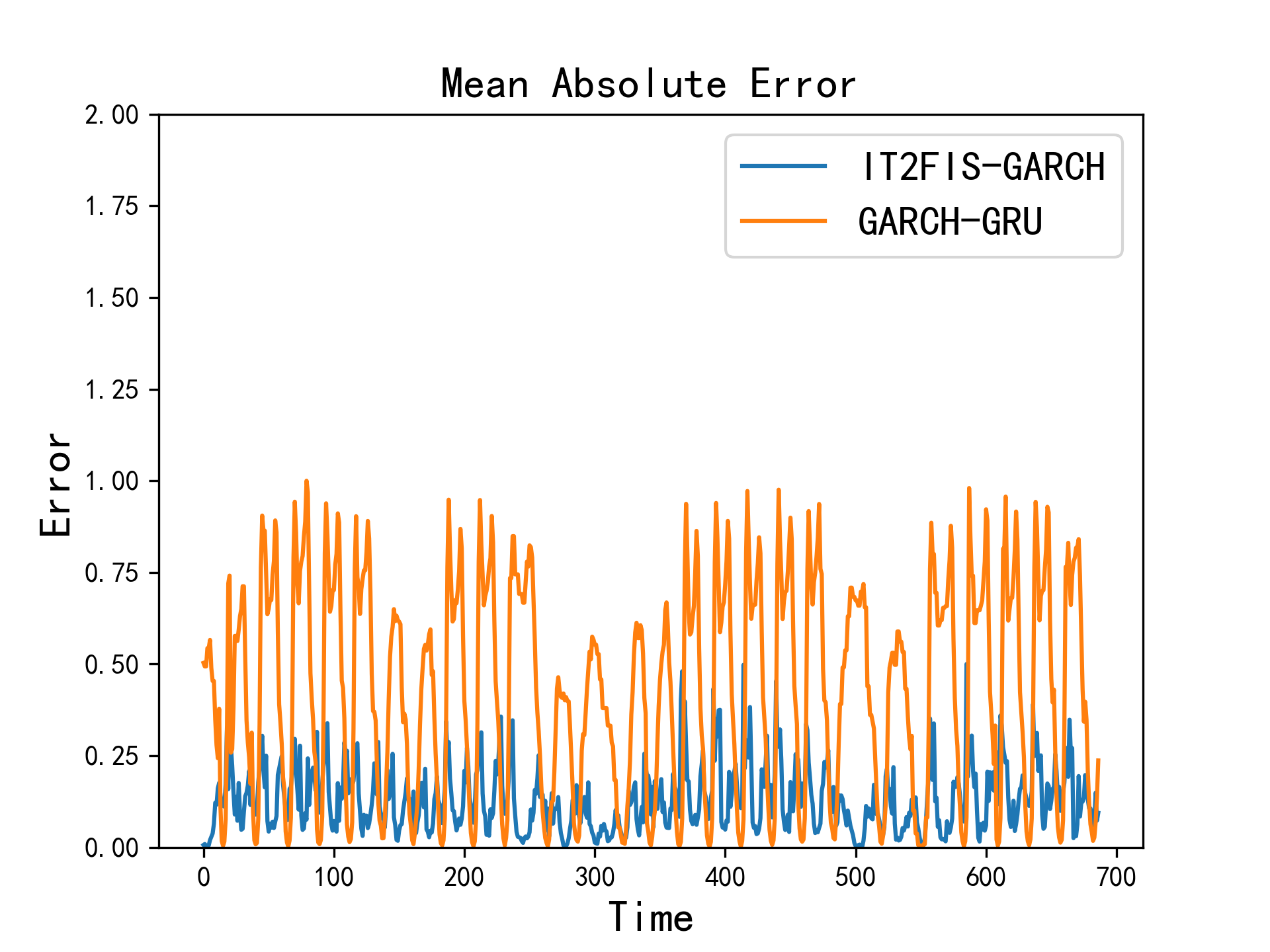}
		\caption{MAE of the Prediction for the 'Traffic' Dataset}
		\label{fig:GRU-traffic}
	\end{subfigure}
	\caption{Comparison of IT2FIS-GARCH model and GARCH-GRU model for prediction in different datasets: (a) "Death", (b) "Traffic". The blue line represents the IT2FIS-GARCH model, and the orange line represents the GARCH-GRU model.}
	\label{fig:comparison2}
\end{figure}

\begin{table}[H]
	\centering
	\tiny
	\caption{Prediction accuracy of IT2FIS-GARCH model and different comparison models on ten datasets (MAE as evaluation metric)}
	\label{tab:MAE}
	\begin{tabular}{ c c c c c c }
		\hline
		Model & IT2FIS  -GARCH & \makecell{Fixed Variance- \\ IT2FIS} & LSTM-IT2FIS & GARCH-GRU & GARCH-TSK \\
		\hline
		AirQualityUCI& 0.15065 & \textbf{0.10697} & 0.34063 & 0.24224 & 0.32851\\
		\hline
		\makecell{Average\_Daily\_ \\ Traffic\_Counts}& \textbf{0.01646} &0.49932&0.49350&0.18384&0.42169 \\
		\hline
		Data& \textbf{0.12703} &0.60887&0.76126&0.65417& 0.39024\\
		\hline
		Death& 0.23945 &\textbf{0.10257}&0.25881&0.26740&0.21966 \\
		\hline
		diabetesl\_2016& \textbf{0.02020} &0.02966&0.36431&0.33559&0.44496 \\
		\hline
		energydata\_complete& \textbf{0.06100} & 0.90287 &0.67432&0.56360& 0.09990\\
		\hline
		ex\_20& 0.52616 & 0.55458&0.56835&\textbf{0.28956}&0.44541 \\
		\hline
		\makecell{household\_power\_ \\ consumption}& \textbf{0.06468} & 0.27369&0.26210&0.35749& 0.68346\\
		\hline
		PRSA\_Data\_Dongsi& 0.13497 &0.26905 &\textbf{0.08420}&0.16755&0.23081 \\
		\hline
		Traffic& \textbf{0.12905} &0.43354 &0.28879&0.43857& 0.19864\\
		\hline
	\end{tabular}
	
\end{table}
\begin{table}[H]
	\centering
	\tiny
	\caption{Prediction accuracy of IT2FIS-GARCH model and different comparison models on ten datasets (MAPE as evaluation metric, \%)}
	\label{tab:MAPE}
	\begin{tabular}{ c c c c c c }
		\hline
		Model & IT2FIS-GARCH & \makecell{Fixed Variance- \\ IT2FIS} & LSTM-IT2FIS & GARCH-GRU & GARCH-TSK\\
		\hline
		AirQualityUCI& 64.11 & 71.25 & 55.73 & 57.73 & \textbf{44.58} \\
		\hline
		\makecell{Average\_Daily\_ \\ Traffic\_Counts}& 73.69 & 50.07&-&69.94& \textbf{18.05} \\
		\hline
		Data& \textbf{31.19} & 57.30&55.52&89.71&31.87 \\
		\hline
		Death& \textbf{2.86} &9.58&83.69&54.05& 4.25\\
		\hline
		diabetesl\_2016& 72.10 &69.13&76.59&87.04&\textbf{42.18} \\
		\hline
		energydata\_complete& \textbf{3.28} & 70.45&38.52&68.00& 6.41\\
		\hline
		ex\_20& \textbf{17.37} &49.96 &99.04&55.65&50.01 \\
		\hline
		\makecell{household\_power\_ \\ consumption}& 22.13 &31.89 &\textbf{19.95}&41.47&57.74 \\
		\hline
		PRSA\_Data\_Dongsi& 14.43 &\textbf{10.95} &64.87&85.59& 13.84\\
		\hline
		Traffic& \textbf{11.23} & 20.66&28.29&62.06&28.70 \\
		\hline
	\end{tabular}
	
\end{table}
Taking the IT2FIS-GARCH model and GARCH-GRU model as examples, it can be seen from Figure \ref{fig:GRU-Death} and Figure \ref{fig:GRU-traffic} that the model proposed in this paper has a smaller fluctuation range of prediction error and makes more accurate predictions. As shown in Table \ref{tab:MAE}, the model proposed in this paper exhibits relatively small mean square errors on six datasets. As shown in Table \ref{tab:MAPE}, the IT2FIS-GARCH model achieved lower MAPE values on five datasets compared to other benchmark models, followed by the GARCH-TSK model, which demonstrated superior error values on three datasets relative to the comparison models.
\subsection{Comparison of $R ^ 2$ experimental results}
\label{$R ^ 2 $}
In this section, the predictive ability of the IT2FIS-GARCH model is discussed based on the existing algorithm ($R ^ 2 $). By calculating $R ^ 2 $, it is possible to determine whether the regression model can effectively explain the changes in the dependent variable, and thus select the optimal regression model.

\begin{table}[H]
	\centering
	\tiny
	\caption{The prediction accuracy of IT2FIS-GARCH model and different comparison models on ten datasets (using $R ^ 2 $ as the evaluation metric)}
	\label{tab:R2}
	\begin{tabular}{ c c c c c c }
		\hline
		Model & IT2FIS-GARCH & \makecell{Fixed Variance- \\ IT2FIS} & LSTM-IT2FIS & GARCH-GRU & GARCH-TSK \\
		\hline
		AirQualityUCI& \textbf{0.942} & 0.783 & 0.664 & 0.279 & 0.666\\
		\hline
		\makecell{Average\_Daily\_ \\ Traffic\_Counts}& \textbf{0.688} &0.682&0.429&0.232& 0.399\\
		\hline
		Data& \textbf{0.899} &0.466&0.889&0.159& 0.891\\
		\hline
		Death& \textbf{0.975} &0.951&0.916&0.609& 0.933 \\
		\hline
		diabetesl\_2016& 0.177 &0.185&0.131&0.370& \textbf{0.432}\\
		\hline
		energydata\_complete& \textbf{0.999} &0.551 &0.895&0.885& 0.977 \\
		\hline
		ex\_20& 0.334 & \textbf{0.458}&0.048&0.295& 0.110\\
		\hline
		\makecell{household\_power\_ \\ consumption}& \textbf{0.192} &0.034 &0.183&0.109& 0.024\\
		\hline
		PRSA\_Data\_Dongsi& 0.818 &0.652 &\textbf{0.960}&0.292&0.163 \\
		\hline
		Traffic& \textbf{0.939} &0.348 &0.394&0.185& 0.746\\
		\hline
	\end{tabular}
	
\end{table}

$R ^ 2 $ is a numerical value between 0 and 1 that measures the model's ability to interpret data. Taking the second dataset as an example, the fixed variance interval type-2 fuzzy inference system has the highest $R ^ 2 $ value, but also a high MAPE value. This means that the model overfits and captures non realistic trends. This phenomenon can cause the model to perform well on training data but poorly on new data, resulting in significant prediction errors According to Table \ref{tab:R2}, it can be observed that the IT2FIS-GARCH model has the highest $R ^ 2 $ value on the six datasets. Based on Table \ref{tab:MAPE}, it can be concluded that the IT2FIS-GARCH model has a higher degree of fit and lower error values on the five datasets. This indicates that the IT2FIS-GARCH model has better predictive performance than other models on most heteroscedastic time series datasets.

\subsection{Experimental results}
\label{Experimental Results}
According to the chart data in Subsection \ref{MSE}, it can be seen that the fixed variance time series model has certain limitations in predicting GARCH-type time series, and the IT2FIS-GARCH model is less affected in the prediction process. According to the chart data in Subsection \ref{RMSE}, it can be seen that the prediction accuracy of the IT2FIS-GARCH model is generally higher than other prediction models on most heteroskedastic time series datasets. Through comparison, it was found that when dealing with the same GARCH-type time series dataset, our model can more effectively capture and adapt to heteroskedasticity in time series by introducing heteroskedasticity features. This highlights the advantage of incorporating heteroskedasticity as a dynamic feature in interval type-2 fuzzy inference systems, which can significantly improve the predictive ability and robustness of the model when processing GARCH-type time series data.

The experimental comparison results in Subsection \ref{Comparison of MAE and MAPE Experimental Results} and Subsection \ref{$R ^ 2 $} indicate that the IT2FIS-GARCH model has a unique "interval" characteristic, which enables it to more comprehensively consider the heteroskedasticity and uncertainty of GARCH-type time series, provide more accurate evaluation basis in the decision-making process, and generate relatively reliable outputs. This model can more accurately describe and handle the degree and range of uncertainty by interval processing the membership degrees of input and output variables, effectively capturing and modeling heteroskedasticity features in time series, thereby improving prediction accuracy and model adaptability. This fully demonstrates the advantages of the IT2FIS-GARCH model in dealing with heteroskedasticity in time series data.

\section{Conclusion}
\label{Conclusion} 
In this innovative study, we propose a variable parameter interval type-2 fuzzy inference system (IT2FIS-GARCH model) that dynamically embeds heteroskedasticity features. The IT2FIS-GARCH model cleverly combines the advantages of interval type-2 fuzzy inference systems and generalized autoregressive conditional heteroskedasticity (GARCH) models. By synergizing the uncertainty-handling capabilities of interval type-2 fuzzy systems with the volatility modeling rigor of generalized autoregressive conditional heteroskedasticity (GARCH), the model achieves two key innovations:

1. Volatility-Driven Interval Adaptation: This article uses the GARCH model to provide interval variance for the Gaussian membership function of  IT2FIS, enabling dynamic adjustment of fuzzy set intervals to align with time-varying volatility regimes.

2. Mean-Variance Coevolution: The defuzzification mechanism based on the IT2FIS-GARCH model optimizes the conditional mean prediction of the GARCH model, jointly optimizing trend and forecasts.

In the experimental phase, we applied the proposed model to multiple datasets for in-depth predictive analysis. Experiments have shown that the IT2FIS-GARCH model exhibits smaller prediction errors compared to other prediction models on multiple datasets. This achievement not only validates the model's ability to handle complex heteroscedastic time series data, but also provides new ideas and methods for future research in the field of time series prediction.

\section*{CRediT Authorship Contribution Statement}
Hongpei Shao: Conceptualization, Dataset, Methodology, Software, Validation, Writing– original draft. Da-Qing zhang: Conceptualization, Supervision, Writing - Review \& Editing. Feilong Lu: Conceptualization, Writing - Review \& Editing.

\section*{Declaration of Competing Interest}
The authors declare that they have no known competing finan cial interests or personal relationships that could have appeared to influence the work reported in this paper.

\section*{Data Availability Statement}
The data supporting this study are openly available in the UCI Machine Learning Repository, Data.gov and Figshare. A complete list of datasets, including their access links, as follows:
\begin{table}[H]
	\centering
	\small 
	\begin{tabularx}{\linewidth}{YYY}
		\hline
		\textbf{Dataset} & \textbf{Source} & \textbf{Identifier} \\
		\hline
		\url{AirQualityUCI} & UCI Machine Learning Repository & DOI: \url{10.24432/C59K5F} \\
		\hline
		\url{energydata_complete} & UCI Machine Learning Repository & DOI: \url{10.24432/C5VC8G} \\
		\hline
		\url{ex_20} & UCI Machine Learning Repository & DOI: \url{10.24432/C58S4T} \\
		\hline
		\url{household_power_ consumption} & UCI Machine Learning Repository & DOI: \url{10.24432/C58K54} \\
		\hline
		\url{PRSA_Data_Dongsi} & UCI Machine Learning Repository & DOI: \url{10.24432/C5RK5G} \\
		\hline
		\url{Traffic} & UCI Machine Learning Repository & DOI: \url{10.24432/C5X60B} \\
		\hline
		\url{Average_Daily_ Traffic_Counts} & Data.gov & \url{https://catalog.data.gov/dataset/average-daily-traffic-counts} \\
		\hline
		\url{Data} & Figshare & \url{https://figshare.com/articles/dataset/Modelling_time-varying_volatility_using_GARCH_models/20681203/2} \\
		\hline
		\url{Death} & Data.gov & \url{https://catalog.data.gov/dataset/death-rates-for-suicide-by-sex-race-hispanic-origin-and-age-united-states-020c1} \\
		\hline
		\url{diabetesl_2016} & Data.gov & \url{https://catalog.data.gov/dataset/diabetes} \\
		\hline	
	\end{tabularx}
\end{table}

\appendix

\end{document}